\documentclass[a4paper,11pt]{article}
\pdfoutput=1 % if you are submitting a pdflatex (i.e. if you have
\usepackage{jheppub} % for details on the use of the package, please
\usepackage{booktabs,makecell, bm}
\usepackage[font=small]{caption}
\usepackage{framed}
\usepackage{hyperref}
\usepackage[export]{adjustbox}
\usepackage{amsmath, tensor}
\DeclareMathOperator{\Tr}{Tr}
\usepackage{physics}
\usepackage{subcaption}
\usepackage{multirow}
\usepackage{adjustbox}

\usepackage{float}

\def\CN{\mathcal{N}}
\newcommand{\D}[1]{\dot{#1}}
\newcommand{\C}[1]{\mathcal{#1}}
\newcommand{\B}[1]{\overline{#1}}

\newcommand{\TR}[1]{\textrm{#1}}

\title{Fortuity and relevant deformation}

\author[a,c]{Jaehyeok Choi,}
\author[b]{Seunggyu Kim}
\affiliation[a]{Department of Physics and Astronomy \& Center for
Theoretical Physics,\\
Seoul National University, Seoul 08826, Korea.}
\affiliation[b]{Department of Physics, Korea Advanced Institute of Science and Technology\\291 Daehak-ro, Yuseong-gu, Daejeon 34141, Republic of Korea.}
\affiliation[c]{School of Physics, Korea Institute for Advanced Study,\\
85 Hoegi-ro, Dongdaemun-gu, Seoul 02455, Republic of Korea}

\emailAdd{jaehyeokchoi@kias.re.kr, sgkim01@kaist.ac.kr}
\emailAdd{}

\abstract{
We investigate the supercharge cohomology of an $\mathcal{N}=1$ relevant deformation of $\mathcal{N}=4$ super Yang-Mills.
By introducing a field redefinition, we integrate out massive fields in a cohomological sense.
Then, we construct the monotone cohomologies corresponding to the Kaluza-Klein particles of the dual supergravity solution.
Some of the monotone cohomologies obey stringy exclusion principle analogous to that of $AdS_3$.
Relatedly, they vanish on the diagonal field configurations, unlike $\mathcal{N}=4$ monotone cohomologies.
We also construct infinitely many fortuitous cohomologies for gauge group SU(2).
We find that unlike $\mathcal{N}=4$ fortuitous cohomologies, they can either be non-vanishing or vanishing on the diagonal fields.
By undoing the field redefinition and taking a suitable UV limit, we show that non-vanishing ones reduce to monotone cohomologies of $\mathcal{N}=4$ SYM, while vanishing ones reduce to fortuitous cohomologies of $\mathcal{N}=4$ SYM.
This implies that the fortuity can arise due to the relevant deformation, while monotonicity is not.
}

\makeatletter
\gdef\@fpheader{}
\makeatother

\begin{document} 
\maketitle
%\flushbottom

%%%%%%%%%%%%%%%%%%%%%%%%%%%%%%%%%%%%%%%%%
%%%%%%%%%%%%%%%%%%%%%%%%%%%%%%%%%%%%%%%%%
%%%%%%%%%%%%%%%%%%%%%%%%%%%%%%%%%%%%%%%%%
%%%%%%%%%%%%%%%%%%%%%%%%%%%%%%%%%%%%%%%%%
%%%%%%%%%%%%%%%%%%%%%%%%%%%%%%%%%%%%%%%%%
%%%%%%%%%%%%%%%%%%%%%%%%%%%%%%%%%%%%%%%%%
\section{Introduction}
Identifying black hole microstates is a key challenge in quantum gravity.
% The AdS/CFT correspondence \cite{Maldacena:1997re} provides a non-perturbative framework that allows us to understand black holes in terms of quantum field theory, where we have stronger theoretical control. 
For the microstates of supersymmetric black holes in anti-de Sitter (AdS) space, one can study the supersymmetric operators in the dual superconformal field theory (SCFT) \cite{Maldacena:1997re}.
The supersymmetric operators, corresponding to the supersymmetric states on the sphere, are annihilated by a supercharge $Q$ and its hermitian conjugate (on the sphere) $Q^\dag=S$, satisfying the Bogomol'nyi-Prasad-Sommerfield (BPS) bound.
From the standard Hodge theoretic argument, such BPS operators are in one-to-one correspondence with $Q$-cohomology classes.
The study of this $Q$-cohomology \cite{Grant:2008sk,Chang:2013fba} has been recently revived \cite{Chang:2022mjp,Choi:2022caq}, following the successful explanation (e.g. \cite{Cabo-Bizet:2018ehj,Choi:2018hmj,Benini:2018ywd}) of black hole entropy using the superconformal index \cite{Kinney:2005ej,Romelsberger:2005eg}, and has been further investigated in various directions \cite{Choi:2023znd,Budzik:2023vtr,Chang:2023zqk,Choi:2023vdm,Chang:2024zqi,Chang:2024lxt,deMelloKoch:2024pcs,Chang:2025rqy,Gadde:2025yoa,Chang:2025mqp,Chen:2025sum,Kim:2025vup,Choi:2025bhi,Chang:2025wgo,Gaikwad:2025ugk,Belin:2025hsg}.
The $Q$-cohomology of general SQFT (holomorphic twist) has also been studied, e.g., \cite{Budzik:2022mpd,Budzik:2023xbr,Bomans:2023mkd,Gaiotto:2024gii,Scheinpflug:2024mtn,Bomans:2025klo,Budzik:2025zvu}.

Cohomology classes corresponding to BPS Kaluza-Klein particle states in the dual AdS gravity theory are termed \emph{gravitons}. Since these operators remain $Q$-closed for arbitrary rank $N$, they are also called \emph{monotone} cohomologies \cite{Chang:2024zqi}. In contrast, cohomology classes that are linearly independent of the gravitons are referred to as \emph{black hole} cohomologies (or, more neutrally, as non-gravitons). Because their closure under $Q$ relies crucially on finite $N$ trace relations, they are alternatively named \emph{fortuitous} cohomologies \cite{Chang:2024zqi}.
% This classification of cohomologies suits well with the theories that do not contain baryonic operators, which cannot be rewritten as multi-traces \cite{conifold}. Such theories are said to admit `holographic covering' \cite{Chang:2024zqi}, and we will focus on such cases in this paper.

We study a mass deformation of $\mathcal{N}=4$ super Yang-Mills (SYM) theory with gauge group $SU(N)$. $\mathcal{N}=4$ SYM consists of three chiral multiplets $(X,Y,Z)$ in the adjoint representation and one vector multiplet with superpotential $W=\mathrm{Tr}X[Y,Z]$ when viewed as an $\mathcal{N}=1$ theory. If one deforms the theory by adding a mass term for one of the chiral multiplets,
$$W=\mathrm{Tr}X[Y,Z]+\frac{M}{2}\mathrm{Tr}Z^2\;,$$
the theory flows to an $\mathcal{N}=1$ SCFT with a quartic superpotential $W=-\frac{1}{2M}\mathrm{Tr}[X,Y]^2$, where the superfield $Z$ is integrated out.
The symmetry $SU(4)_R$ is broken to $SU(2)\times U(1)_R$.
This theory is sometimes called the Leigh-Strassler (LS) SCFT \cite{Leigh:1995ep}.
This quartic superpotential is exactly marginal in the IR, while it is irrelevant and not renormalizable in the UV.
% Since the superconformal R-charges of IR adjoint chiral superfields are $1/2$, there is no known simple kinetic term (K\"ahler potential) for chiral multiplets.
Our aim is to investigate the cohomology of this IR SCFT.

The holographic dual of the LS SCFT is type $\rm IIB$ superstring theory in the Pilch-Warner background \cite{Pilch:2000ej}, which is a warped product $AdS_5\times \tilde{S}^5$, where $\tilde{S}^5$ is a squashed five-sphere whose isometry is broken to $SU(2)\times U(1)_R$ from $SU(4)_R$.
The full Kaluza-Klein (KK) particle spectrum of type $\rm IIB$ supergravity in the Pilch-Warner background was obtained rather recently in \cite{Bobev:2020lsk} using the exceptional field theory method. 
Among them, the shortened multiplets of $\mathcal{N}=1$ superconformal symmetry $SU(2,2|1)$ contain states that correspond to the monotone operators in the LS SCFT.
We explicitly find representatives of cohomology classes corresponding to the BPS KK particles. 
The spectrum of shortened multiplets and corresponding monotone cohomologies is summarized in Table \ref{tab: Single-trace graviton}.

Somewhat surprisingly, we find that some towers of single-trace graviton cohomologies are of \emph{non-Coulomb} type.
Here, \emph{Coulomb type cohomology} is defined as a cohomology class whose representatives are non-vanishing when all the fields are diagonal (and non-Coulomb when they are vanishing) \cite{Chang:2025mqp,Choi:2025bhi}. 
Note that this definition is well-defined if the $Q$-action yields a commutator, making the $Q$-exact operators always non-Coulomb.
This also means that any Coulomb-type operator is non-exact.
Non-Coulomb operators can either be exact or non-exact. 
In fact, all the known non-graviton cohomologies in $\C{N}=4$ SYM can be chosen to be of non-Coulomb type by subtracting suitable multi-gravitons \cite{Choi:2023znd,Choi:2023vdm,deMelloKoch:2024pcs,Gadde:2025yoa}.

It has been implicitly assumed in \cite{Choi:2023znd,Choi:2023vdm} that all the graviton cohomologies (even the multi-trace gravitons) are of Coulomb type.
This assumption originates from the computation of the chiral ring partition function in $\mathcal{N}=4$ SYM \cite{Kinney:2005ej}, and the fact that single-trace gravitons are all obtained as super-descendants of chiral primaries. 
This greatly simplifies the procedure for identifying (indicial) charges of non-graviton cohomologies.
This is because determining whether a graviton cohomology (which is always $Q$-closed) is $Q$-exact reduces to checking whether it vanishes on diagonal field configurations.
However, it was realized in \cite{Kim:2025vup,conifold} that in the theories other than $\mathcal{N}=4$ SYM, there are multi-trace graviton cohomologies that are of non-Coulomb type (in a suitably generalized sense). Furthermore, it was found in \cite{Chang:2025mqp} that even in (the anti-chiral ring of) $\C{N}=4$ SYM with gauge group $SO(7)$, there exists a multi-trace graviton cohomology that is non-Coulomb and classically non-exact\footnote{In this case, it was shown in \cite{Choi:2025bhi} that such a non-Coulomb graviton is, in fact, $Q$-exact once a quantum correction due to the Konishi anomaly is taken into consideration. The non-Coulomb graviton and a fortuitous cohomology pair up and get lifted.}.\footnote{
In fact, the existence of non-Coulomb cohomologies in the chiral ring of pure $\C{N}=1$ SYM with gauge group $SO(N)$ (for $N\geq 7$) was already known in \cite{Witten:2003ye}, and the same cohomologies exist in $\C{N}=4$ $SO(N)$ as well. If we define a glueball superfield $S\sim \mathrm{Tr}W_\alpha W^\alpha$, it is nilpotent in the \emph{classical} chiral ring: $S^h=0$, where $h$ is the dual Coxeter number, while $S^{h-1}\neq 0$. This was shown in \cite{Witten:2003ye} for gauge groups $SU(N),SO(N),Sp(N)$. The proof of $S^{h-1}\neq 0$ for the $SO(N)$ case in \cite{Witten:2003ye} was rather indirect, using the instanton calculations in the literature, and later it was proven algebraically in \cite{Etingof:2003fy}.
For $N\geq 7$, the rank $r=\lfloor\frac{N}{2}\rfloor$ is smaller than $h-1=N-3$ for $SO(N)$, making $S^{r+1},S^{r+2},\cdots,S^{h-1}$ non-Coulomb.
}
In our case, we find that there are infinite towers of \emph{single-trace} gravitons that are non-Coulomb.

% It might be related to the fact that the moduli space of vacua of the LS SCFT has a 

Also, we find that one tower of such non-Coulomb single trace gravitons (which we call $w_{n}$, where $n=0,1,2,\cdots$) becomes $Q$-exact when its IR superconformal R-charge (absolute value) is too large.
More precisely, $w_n$ has R-charge $-(n+1)$, and $w_n$ is $Q$-exact when $n\geq N-1$ for gauge group $SU(N)$.
This is analogous to the stringy exclusion principle in $AdS_3$ \cite{Maldacena:1998bw}, in the sense that there is an upper bound on the absolute value of R-charge. 
Unlike in the $\C{N}=4$ SYM case, the fast-rotating (single) particle corresponding to $w_{n\geq N-1}$ in the internal direction of the gravity dual does not `polarize' into a D3-brane (giant graviton \cite{McGreevy:2000cw,Grisaru:2000zn,Hashimoto:2000zp}) due to the RR-flux.
Instead, that particle becomes non-BPS for large angular momentum in the internal direction.
On the other hand, other particles corresponding to Coulomb type cohomologies behave similarly to the $\C{N}=4$ gravitons.

The existence of non-Coulomb single-trace gravitons is closely related to the fact that the (classical) $Q$-action of the mass-deformed $\C{N}=4$ SYM, which we use to define our cohomology problem, is not a commutator.
The supersymmetry action on a chiralino related to $Z$ is deformed from $[X,Y]$ to $[X,Y]+MZ$.
We will introduce a `change of variables' (or field redefinitions) that `integrates out' massive fields in a cohomological sense.
This will be explained in Sections \ref{sub:change of variables} and \ref{sub: integrating out}.
% We show that one can always choose a representative of a cohomology class without using the massive fields in the IR coordinates.
After this change of variables, the $Q$-action on the fields becomes commutators.
For example, the $Q$-actions on chiralinos are given by the derivatives of the quartic IR superpotential.
The non-Coulomb single-trace gravitons are non-Coulomb only \emph{after} the change of variables.
\emph{Before} the change of variables, they can be represented as Coulomb-type \emph{operators}. An explicit demonstration for $w_0$ is in Section \ref{section:IR cohomology in UV coordinates}.

As a mnemonic, we refer to the original $\C{N}=4$ SYM fields as `UV coordinate' and the fields after the change of variables as `IR coordinate', even though we are still dealing with classical cohomology near UV.
In this work, we use the IR coordinate to find cohomology representatives and then express them in the UV coordinate in Section \ref{section:IR cohomology in UV coordinates}.
This is because, in the IR coordinate, the massive fields can be integrated out.
More precisely, we show in Section \ref{sub: integrating out} that one can always choose a representative of a cohomology class without using the massive fields in the IR coordinates.
This fact is not obvious in the UV coordinates.
% We regard this as a definition of the classical cohomology problem for the IR SCFT.

Note that when the $Q$-action does not yield commutators, the Coulomb operators are not necessarily non-exact. 
In fact, we will see that part of the single-trace gravitons of $\C{N}=4$ SYM (which are Coulomb in the UV coordinate) becomes exact after the mass-deformation in Section \ref{section:IR cohomology in UV coordinates}.

Specializing to gauge group $SU(2)$, we explicitly construct infinitely many (conformal primary) non-graviton operators in the IR coordinates.
Following the strategy of \cite{Choi:2023znd,Choi:2023vdm}, we first count the Coulomb gravitons and compute their index. By subtracting the Coulomb graviton index from the full index, we obtain information about non-gravitons and non-Coulomb gravitons.
It turns out that the non-gravitons start to appear at a much earlier stage compared to $\C{N}=4$ SYM with gauge group $SU(2)$ \cite{Chang:2022mjp,Choi:2022caq,Choi:2023znd}.
We define the index with fugacity $t$ as
$$\mathrm{Tr}(-1)^F t^{3(2j-r)},$$
where $j$ is the Cartan of the left factor of $SU(2)\times SU(2)$, and $r$ is the superconformal R-charge (different for UV and IR).
The convention will be explained in detail in Section \ref{sub:index}. 
This index receives contributions from operators that are annihilated by $Q\equiv Q_-$ and $S^-=(Q_-)^\dag$, which include the anti-chiral ring.
Then, the non-gravitons first appear as $4t^{21/2}$ in the $SU(2)$ LS SCFT, while a non-graviton first appears as $-t^{24}$ in $SU(2)$ $\C{N}=4$ SYM.

Unlike the $\mathcal{N}=4$ SYM non-graviton cohomologies \cite{Choi:2023znd,Choi:2023vdm,deMelloKoch:2024pcs,Gadde:2025yoa}, we find that the non-gravitons in the LS SCFT can be of Coulomb type. 
That is to say, there are no graviton cohomologies that can render such non-gravitons non-Coulomb by subtracting them.
In fact, until order $-t^{24}$, all the non-graviton cohomologies are of Coulomb type.
Even though they are fortuitous, i.e. $Q$-closed only after using finite $N$ trace relations, and they are independent of gravitons, we show that in the small mass limit (UV limit), they reduce to (multi) graviton cohomologies of $\C{N}=4$ SYM in Section \ref{section:IR cohomology in UV coordinates} by translating them into UV coordinates.
In other words, some of the multi-gravitons of $\C{N}=4$ SYM become non-gravitons after the relevant deformation.
The `fortuity' of these formerly monotone cohomologies arises due to the relevant deformation.

At $-t^{24}$, we explicitly construct the first non-graviton operator that is non-Coulomb, which we denote as $O_{16}$ in Section \ref{sub:nonCoulombnongraviton}. 
% Since the charges (under $U(1)_R$, flavor $SU(2)$, and angular momenta) match those of the lightest non-graviton of $SU(2)$ $\C{N}=4$ SYM, this non-Coulomb non-graviton must reduce to the lightest non-graviton of $SU(2)$ $\C{N}=4$ SYM in the small mass limit in the UV coordinates.
It has Lorentz spin $j=5/2$ and IR R-charge $r_{IR}=-3$.
On the other hand, the lightest non-graviton operator (`$O_0$' in \cite{Choi:2023znd}) in $SU(2)$ $\C{N}=4$ SYM has spin $j=5/2$ and UV R-charge $r_{UV}=-3$. 
They are both singlets under the unbroken $SU(2)$ global symmetry.
% We find that its charges (under $U(1)_R$, the flavor $SU(2)$, and the angular momenta) coincide with those of the lightest non-graviton operator (`$O_0$' in \cite{Choi:2023znd}) in $SU(2)$ $\C{N}=4$ SYM. 
We show in section \ref{section:IR cohomology in UV coordinates} that $O_{16}$ reduces to `$O_0$' in the $M\to 0$ limit. However, it is highly non-trivial, unlike the Coulomb non-graviton cases.

The paper is structured as follows. In Section~\ref{section 2}, we briefly review the IR SCFT obtained via an RG flow starting from mass-deformed $\C{N}=4$ SYM, and compute the superconformal indices for both UV and IR theories. 
In Section \ref{section 3}, we discuss the $Q$-action and its cohomology for the IR SCFT, whose Lagrangian is not known.
Starting with the UV Lagrangian, we introduce a change of variables mentioned earlier in Section \ref{sub:change of variables} and explain how it integrates out the massive degrees of freedom cohomologically in Section \ref{sub: integrating out}. 
% We regard this as a definition of the classical cohomology problem for the IR SCFT.
We then address the classification of cohomologies into Coulomb and non-Coulomb types and discuss their relation to the moduli space of vacua in \ref{sub:coulomb noncoulomb}.
% We also discuss the BMN truncation, which is a consistent truncation of the cohomology problem at the classical level.
In Section \ref{section 4}, we present our main results on the black hole cohomology problem in the LS SCFT.
We explicitly construct graviton cohomologies in the LS SCFT and illustrate how to obtain them with the example $w_n$ in Section \ref{sub:gravitons}.
We then compute the $SU(2)$ BMN Coulomb graviton index analytically and subtract it from the full $SU(2)$ BMN index in Section \ref{sub:coulomb grav index}.
We construct infinitely many non-graviton cohomologies that are of Coulomb type in \ref{sub: Coulomb nongrav} and one non-Coulomb non-graviton cohomology in \ref{sub:nonCoulombnongraviton}.
In Section \ref{section:IR cohomology in UV coordinates}, we express the `IR' cohomologies in the UV coordinates and identify the $\C{N}=4$ origins.
In Section \ref{section: discussion}, we conclude with remarks.
% In Appendix \ref{apndx:hodge}, we briefly review the standard Hodge theory argument.
In Appendix \ref{minimal check}, we explicitly demonstrate that $w_n$ are $Q$-exact for $n\geq 1$ in $SU(2)$ theory.
In Appendix \ref{appendix: O16UV}, we show the explicit expression of non-Coulomb non-graviton operator in the UV coordinates.
% In Appendix \ref{freebpspartition}, we compute the free BPS partition function that helped us find cohomologies or check non-exactness.

% In Section~\ref{section 3}, we address the framework for defining supercharge cohomology using the UV (mass-deformed) Lagrangian and argue that this approach is conceptually consistent for studying the IR theory. We also discuss the BMN truncation, a consistent subsector used in our analysis. 
% In Section~\ref{section 4}, we present our main results: the identification of graviton operators in one-to-one correspondence with the Kaluza–Klein spectrum on the dual gravity side, and the classification and counting of graviton and non-graviton cohomologies using the eigenvalue method. We explicitly construct several examples of non-graviton operators and show that, and show that, unlike in $\C{N}=4$ SYM, non-eigenvalue gravitons also exist.

%%%%%%%%%%%%%%%%%%%%%%%%%%%%%%%%%%%%%%%%%
%%%%%%%%%%%%%%%%%%%%%%%%%%%%%%%%%%%%%%%%%
%%%%%%%%%%%%%%%%%%%%%%%%%%%%%%%%%%%%%%%%%
%%%%%%%%%%%%%%%%%%%%%%%%%%%%%%%%%%%%%%%%%
%%%%%%%%%%%%%%%%%%%%%%%%%%%%%%%%%%%%%%%%%
%%%%%%%%%%%%%%%%%%%%%%%%%%%%%%%%%%%%%%%%%
\section{Relevant deformation of $\mathcal{N}=4$ SYM theory}\label{section 2}
%Leigh and Strassler \cite{Leigh:1995ep} studied 
The most general renormalizable deformation of $\CN=4$ SYM, preserving (minimal) supersymmetry, is given by the superpotential 
\begin{equation}
    \Delta W=\frac{1}{6}h^{ijk}\Tr(\Phi_i\Phi_j\Phi_k)+\frac{1}{2}m^{ij}\Tr(\Phi_i\Phi_j) \ ,
\end{equation}
where $h^{ijk}$ is a marginal coupling that is totally symmetric in $(i,j,k)$, and $m^{ij}$ is a symmetric mass matrix.
The marginal couplings $h_{ijk}$ in general break $SU(3)$ flavor symmetry (in the sense of $\mathcal{N}=1$ theory). There are a total of 11 marginal operators in $\CN=4$ SYM theory (gauge coupling + $h_{ijk}$'s). At a generic point in the conformal manifold, 8 of them will combine with the broken $SU(3)$ symmetry generators to become marginally irrelevant. Therefore, there remain 3 exactly marginal couplings spanning the three-dimensional conformal manifold \cite{Leigh:1995ep, Green:2010da}. Note that superpotentials are generally non-BPS.
% (\textbf{JC: superpotentials are usually non-BPS. e.g. $\C{N}=4$ SYM superpotential.}) 

The only (supersymmetric) relevant operators are given by the quadratic Casimirs of the form $\Tr \Phi_i^2$, which are the mass terms for the adjoint chiral multiplets. We give a mass to one of them
\begin{equation}
    \Delta W=\frac{M}{2}\Tr(\Phi_3^2)\ .
\end{equation}
Integrating out $\Phi_3$ yields, in the IR, an $\C{N}=1$ fixed point with two massless adjoint chiral fields and a quartic superpotential:
\begin{equation}
    W_{\C{N}=4}^{\text{Mass}}=\Tr(\Phi_3[\Phi_1,\Phi_2])+\frac{M}{2}\Tr\Phi_3^2 \;\to\; W_{IR}=-\frac{1}{M}\Tr([\Phi_1,\Phi_2]^2) \ . 
\end{equation}
The quartic superpotential $W_{IR}$ in the infrared becomes a marginal operator. 
Note also that the superpotential preserves $SU(2)_F$ symmetry. A modern review of this IR SCFT (and its gravity dual) can be found in \cite{Bobev:2025gzu}.

% (\textbf{JC: 
% % e.g. \cite{Aharony:1999ti} page 140 
% $\Lambda_{YM}\sim M e^{-c/g_{YM}^2(M)N}$ - general $\C{N}=1$ 1-loop running. $c$ is independent of $N$. $\Lambda_{YM}/M\ll 1$ means small 't Hooft coupling. Near UV cohomology with small $g_{YM}^2N$ is $\Lambda_{YM}/M\ll 1$ cohomology? which could be same(?) as IR cohomology?.})

% (\textbf{JC: Maybe could argue that the cohomology is independent of $M$ except for $M=0$ using Witten's conjugation trick?\cite{Witten:1982df} i.e. $Q\to P^{-1}QP$ where $P=\exp(\int_{S^3}\mathrm{Re}(\frac{\Delta M}{2}\Phi_3^2))$—if $P$ is a well-defined operator, conjugation by $P$ changes the value of $M$ and the BPS state space remains isomorphic. One caveat is - once we mass-deform the theory, there is no proper state-operator correspondence. Another problem is the renormalization of D-term which can't be implemented using the conjugation trick.})

The IR fixed point lies on a particular $SU(2)\subset SU(3)$ invariant locus of the full conformal manifold, obtained by imposing the exact marginality conditions on all couplings $h^{ijk}$ and $m^{ij}$. In particular, the conformal manifold is three-dimensional and is believed to have no weakly-coupled point, $i.e.$ it is compact.
% \footnote{A conformal manifold is a family of CFTs connected by exactly marginal deformations. It is naturally equipped with the Zamolodchikov metric, defined through the two-point functions of marginal operators. For example, in $\C{N}=4$ SYM with the complexified gauge coupling $\tau=\frac{\theta}{2\pi}+i\frac{4\pi}{g^2}$, the metric takes the exact form $g_{\tau\B{\tau}}\sim(\Im\tau)^{-2}$. In the free limit $g\to0$ ($\Im\tau\to\infty$), the geodesic distance diverges logarithmically, $d\sim\int^\infty\sqrt{g_{\tau\B{\tau}}}|d\tau|\sim\int^\infty\frac{d(\Im\tau)}{\Im\tau}$. Thus the free point lies at infinite distance in the Zamolodchikov metric. In this sense, the conformal manifold of $\C{N}=4$ SYM is non-compact. For further discussions, see \cite{Asnin:2009xx, Green:2010da, Buican:2014sfa, Gerchkovitz:2014gta}.}
Consequently, the IR SCFT are intrinsically strongly coupled, which makes the definition of the cohomology problem challenging. Nevertheless, in Section~\ref{section 3} we will adopt a prescription analogous to that of \cite{Romelsberger:2005eg,Romelsberger:2007ec}, commonly used in the supersymmetry context under RG flow and in computations of the superconformal index \cite{Romelsberger:2005eg,Kinney:2005ej,Gadde:2010en}, in order to properly define the supercharge cohomology problem for the IR SCFT.

The gravity dual of the mass deformation is a kink solution of five-dimensional $\mathcal{N}=8$ $SO(6)$ gauged supergravity that interpolates between two supersymmetric $AdS_5$ vacua, corresponding to the $\mathcal{N}=4$ and $\mathcal{N}=1$ IR fixed points, respectively \cite{Khavaev:1998fb,Freedman:1999gp}. The ten-dimensional solution dual to the IR SCFT is then obtained by uplifting the IR $AdS_5$ vacuum to a solution of type IIB supergravity \cite{Pilch:2000ej}, which takes the form of a warped $AdS_5\times\Tilde{S}^5$. The internal space $\tilde{S}^5$ is a deformed $S^5$, induced by non-trivial two-form fluxes, and preserves an $SU(2)\times U(1)$ isometry. The full ten-dimensional mass-deformed RG flow solution is also constructed in \cite{Pilch:2000fu}.

\subsection{Superconformal index} \label{sub:index}
For later use, we first compute the index for both the $\C{N}=4$ SYM theory and Leigh-Strassler (LS) SCFT. In general, an $\C{N}=1$ SCFT has the superconformal symmetry $SU(2,2|1)$, whose bosonic subgroup is
\begin{equation}
    SO(4,2)\times U(1)_R
\end{equation}
with supercharges
\begin{equation}
    Q_\alpha,\quad \B{Q}_{\D{\alpha}},\quad S^\alpha=(Q_\alpha)^\dagger,\quad \B{S}^{\D{\alpha}}=(\B{Q}_{\D{\alpha}})^\dagger
\end{equation}
The most relevant anti-commutator relations for our purposes are
\begin{equation}
    \begin{split}
        2\{Q_\alpha,Q^{\dagger \beta}\}=\delta_\alpha^\beta\left(\Delta+\frac{3}{2}r\right)+2\tensor{J}{_\alpha^\beta}\\
        2\{\B{Q}_{\D{\alpha}},\B{Q}^{\dagger \D{\beta}}\}=\delta_{\D{\alpha}}^{\D{\beta}}\left(\Delta-\frac{3}{2}r\right)+2\tensor{\B{J}}{_{\D{\alpha}}^{\D{\beta}}}
    \end{split}
\end{equation}
where $\Delta$ is the conformal dimension, $\tensor{J}{_\alpha^\beta}$ and $\tensor{\B{J}}{_{\D{\alpha}}^{\D{\beta}}}$ are the generators of $SU(2)\times \B{SU(2)}$ with eigenvalues $j$ and $\B{j}$, respectively, and $r$ is the superconformal $U(1)_R$ charge. In particular, we are interested in the states that are annihilated by $Q\equiv Q_-$ and $S=S^-=Q^\dagger$, which satisfy
\begin{equation}\label{eqn: BPS inequality}
    2\{Q,Q^\dagger\}=\Delta+2\tensor{J}{_-^-}+\frac{3}{2}r=\Delta-2j+\frac{3}{2}r\geq 0
\end{equation}
The index for a generic $\C{N}=1$ SCFT, which counts the states that saturate the inequality \eqref{eqn: BPS inequality}, is defined as
\begin{equation}
    \C{I}(\Delta,j,\B{j})=\Tr(-1)^F e^{-\frac{\beta}{2}\delta_-} t^{2(\Delta+j)}y^{2\B{j}}
\end{equation}
Note that the maximal commutant of $Q$ is $SU(2,1)\subset SU(2,2|1)$, which has rank two and is generated by $2(\Delta+j)$ and $2\B{j}$. See also the Table~\ref{tab: N=1 SCFT}.

\begin{table}[t]
\renewcommand{\arraystretch}{1.5}
\centering
\begin{tabular}{|c|c|c|c|c|c|}
\hline
 $Q$ &$\Delta$ & $SU(2)$ & $\B{SU(2)}$ & $U(1)_R$ & $2\{Q,Q^\dagger\}$ \\ \hline\hline
 $Q_{-}$& $\frac{1}{2}$ & $-\frac{1}{2}$ & $0$ & $-1$ & $\delta_-=\Delta-2j_1+\frac{3}{2}r$ \\ \hline
 $Q_{+}$& $\frac{1}{2}$ & $\frac{1}{2}$ & $0$ & $-1$ &  $\delta_+=\Delta+2j_1+\frac{3}{2}r$ \\ \hline\hline
 $\B{Q}_{\D{-}}$& $\frac{1}{2}$ & $0$ & $-\frac{1}{2}$ & $1$ & $\delta_{\D{-}}=\Delta-2j_2-\frac{3}{2}r$ \\ \hline
 $\B{Q}_{\D{+}}$& $\frac{1}{2}$ & $0$ & $+\frac{1}{2}$ & $1$ & $\delta_{\D{+}}=\Delta+2j_2-\frac{3}{2}r$ \\ \hline
\end{tabular}
\caption{\label{tab: N=1 SCFT}In our conventions, $Q_\alpha$ has $r = -1$, and $\B{Q}_{\D{\alpha}}$ has $r = 1$. The hermitian conjugation flips the sign of $r$.}
\end{table}

Let us begin with $\C{N}=4$ SYM theory with a single vector multiplet. This multiplet can be decomposed into one $\C{N}=1$ vector multiplet and three $\C{N}=1$ chiral multiplets:
\begin{equation}
    \begin{split}
        \text{$\C{N}=1$ vector multiplet: }& (A_{\alpha\D{\beta}},\; \lambda_\alpha,\; \B{\lambda}_{\D{\alpha}})\\
        \text{$\C{N}=1$ chiral multiplet: }& (\phi_m,\; \B{\phi}^m,\; \psi_{m\alpha},\; \B{\psi}^m_{\D{\alpha}})
    \end{split}
\end{equation}
Here, $m = 1,2,3$ is the $SU(3)$ index that reflects the breaking of $SU(4)_R$ symmetry to $SU(3) \times U(1)_R$. An upper $SU(3)$ index denotes the fundamental representation, while a
lower index indicates the anti-fundamental representation. In the basis $(T_1,T_2,T_3)$, the Cartan generators of $SU(4)_R$, the $U(1)_R$ superconformal R-symmetry generator reads
\begin{equation}
    -\frac{1}{3}(3T_1+2T_2+T_3)
\end{equation}
The minus sign is due to the convention of choosing $\B{\phi}^m$ to transform in the fundamental representation of $SU(3)$. Another basis, $(R_1, R_2, R_3)$, is sometimes used for the Cartan generators of $SO(6)_R$. These are related by
\begin{equation}
    R_1=T_2+\frac{T_1+T_3}{2},\qquad R_2=\frac{T_1+T_3}{2},\qquad R_3=\frac{T_1-T_3}{2}
\end{equation}
In this basis, the $U(1)_R$ superconformal R-symmetry generator becomes
\begin{equation}
    -\frac{2}{3}(R_1+R_2+R_3)
\end{equation}

Among the 16 supercharges $Q^I_\alpha$ and $\B{Q}_{I\dot{\alpha}}$ in $\C{N} = 4$ SYM, which belong to $PSU(2,2|4)$, we select a specific supercharge, $Q=Q_-^4$, with quantum numbers
\begin{equation}
    (\Delta,j,\B{j},T_1,T_2,T_3)=\left(\frac{1}{2},-\frac{1}{2},0,1,0,0\right)
\end{equation}
or equivalently,
\begin{equation}
    (\Delta,J_1,J_2,R_1,R_2,R_3)=\left(\frac{1}{2},-\frac{1}{2},-\frac{1}{2},\frac{1}{2},\frac{1}{2},\frac{1}{2}\right)
\end{equation}
where $J_1=j+\B{j}$ and $J_2=j-\B{j}$. The BPS inequality \eqref{eqn: BPS inequality} then takes the form
\begin{equation}
    \Delta-(J_1+J_2+R_1+R_2+R_3)\geq0
\end{equation}
and the elementary fields that saturate it are
\begin{equation}
    \B{\phi}^m,\qquad \psi_{m+},\qquad F_{++},\qquad \B{\lambda}_{\D{\alpha}},\qquad \partial_{+\D{\alpha}}
\end{equation}
By multiplying these ``letters'' and contracting gauge indices pairwise, one constructs general gauge-invariant BPS operators in the free theory. These operators contribute to the superconformal index, which counts them in a way that remains invariant under the RG flow and thus independent of the coupling.

We now present the full single-letter index for $\C{N}=4$ SYM theory as calculated in \cite{Kinney:2005ej}:
\begin{equation}
    i_s=\Tr_{\TR{letters}}(-1)^F t^{2(\Delta+j)}u^{2\B{j}}v^{T_2}w^{T_3}=1-\frac{(1-\frac{t^2}{w})(1-\frac{t^2w}{v})(1-t^2v)}{(1-t^3u)(1-\frac{t^3}{u})}
\end{equation}
where the free equations of motion must be imposed as a constraint by subtracting the contributions from $\partial_{+\D{\alpha}}\B{\lambda}^{\D{\alpha}}=0$. In the other basis, we evaluate
\begin{equation}
    i_s=\Tr_{\TR{letters}}(-1)^Fp^{J_1}q^{J_2}x^{R_1}y^{R_2}z^{R_3}=1-\frac{(1-x)(1-y)(1-z)}{(1-p)(1-q)}
\end{equation}
where we introduced a fifth auxiliary fugacity with the additional constraint $pq = xyz$. These two single-letter indices are related through the following fugacity maps:
\begin{equation}
    p=t^3u,\quad q=\frac{t^3}{u},\quad x=t^2v,\quad y=\frac{t^2w}{v},\quad z=\frac{t^2}{w}
\end{equation}

Next, we consider the IR theory, which can be obtained from $\C{N}=4$ SYM by introducing a mass for one of the $\C{N}=1$ adjoint chiral fields and flowing to the fixed point. The UV and IR superconformal $U(1)_R$ charges are given, respectively, by
\begin{equation}\label{eq:ruvrir}
    \begin{split}
        r^{UV}&=-\frac{2}{3}(R_1+R_2+R_3)=-\frac{1}{3}(3T_1+2T_2+T_3)\\
        r^{IR}&=-\frac{1}{2}(R_1+R_2+2R_3)=-\frac{1}{2}(2T_1+T_2)
    \end{split}
\end{equation}
Moreover, the IR SCFT has an $SU(2)_F$ flavor symmetry that rotates the chiral superfields $\Phi_1$ and $\Phi_2$, with a Cartan generator given by
\begin{equation}
    r_F=\frac{1}{2}(R_2-R_1)=-\frac{1}{2}T_2
\end{equation}
The refined index for IR SCFT is given by
\begin{equation}
    \C{I}=\Tr(-1)^F t^{3(2j-r^{IR})}u^{2\B{j}}s^{2r_F}
\end{equation}

From Table~\ref{tab: Listing free BPS letters for LS theory}, which lists the charges of the BPS letters in $\C{N}=4$ SYM and the IR SCFT, we see that the scalar and fermionic components of the adjoint chiral field $\Phi_3$, along with those of its conjugate, contribute to the single-letter index as follows.
\begin{equation}
    \B{\phi}^3:\;t^3u^0s^0,\qquad \psi_{3+}:\;-t^3 u^0 s^0 \ .
\end{equation}
As a result, the $\Phi_3$ chiral multiplet does not contribute to the index, consistent with $\Phi_3$ being integrated out along the RG flow from $\C{N} = 4$ SYM to IR SCFT. From remaining two adjoint chiral multiplet and the vector multiplet, we obtain
\begin{equation}
    i_s^{IR}=\frac{t^{\frac{3}{2}}(x+\frac{1}{x})-t^3(u+\frac{1}{u})-t^{\frac{9}{2}}(x+\frac{1}{x})+2t^6}{(1-t^3u)(1-\frac{t^3}{u})} \ .
\end{equation}
Note that this function can be mapped to $i_s^{UV}$ by substituting
\begin{equation}
    v\;\to\; \frac{s}{\sqrt{t}},\qquad w\;\to\;\frac{1}{t} \ .
\end{equation}
This relation between the UV and IR indices is expected, as the $\C{N}=4$ SYM theory and the IR SCFT are connected by RG flow.

\begin{table}[t]
    \centering
    \begin{tabular}{c|c|c|c|c|c|c|c|c}
        Fields & $(\Delta^{UV},J_1,J_2)$ & $(j,\B{j})$ & $(R_1,R_2,R_3)$ & $(T_1,T_2,T_3)$ & $\Delta^{IR}$ & $r^{UV}$ & $r^{IR}$ & $r_F$ \\\hline\hline
        $\B{\phi}^1$ & $(1,0,0)$ & $(0,0)$ & $(1,0,0)$ & $(0,1,0)$ & $\frac{3}{4}$ & $-\frac{2}{3}$ & $-\frac{1}{2}$ & $-\frac{1}{2}$  \\
        $\B{\phi}^2$ & $(1,0,0)$ & $(0,0)$ & $(0,1,0)$ & $(1,-1,1)$ & $\frac{3}{4}$ & $-\frac{2}{3}$ & $-\frac{1}{2}$ & $\frac{1}{2}$ \\
        $\B{\phi}^3$ & $(1,0,0)$ & $(0,0)$ & $(0,0,1)$ & $(1,0,-1)$ & $\frac{3}{2}$ & $\frac{2}{3}$ & $-1$ & $0$ \\
        $\psi_{1+}$ & $(\frac{3}{2},\frac{1}{2},\frac{1}{2})$ & $(\frac{1}{2},0)$ & $(-\frac{1}{2},\frac{1}{2},\frac{1}{2})$ & $(1,-1,0)$ & $\frac{5}{4}$ & $-\frac{1}{3}$ & $-\frac{1}{2}$ & $\frac{1}{2}$ \\
        $\psi_{2+}$ & $(\frac{3}{2},\frac{1}{2},\frac{1}{2})$ & $(\frac{1}{2},0)$ & $(\frac{1}{2},-\frac{1}{2},\frac{1}{2})$ & $(0,1,-1)$ & $\frac{5}{4}$ & $-\frac{1}{3}$ & $-\frac{1}{2}$ & $-\frac{1}{2}$ \\
        $\psi_{3+}$ & $(\frac{3}{2},\frac{1}{2},\frac{1}{2})$ & $(\frac{1}{2},0)$ & $(\frac{1}{2},\frac{1}{2},-\frac{1}{2})$ & $(0,0,1)$ & $2$ & $-\frac{1}{3}$ & $0$ & $0$ \\\hline\hline
        $F_{++}$& $(2,1,1)$ & $(1,0)$ & $(0,0,0)$ & $(0,0,0)$ & $2$ & $0$ & $0$ & $0$ \\
        $\B{\lambda}_{\dot{+}}$& $(\frac{3}{2},\frac{1}{2},-\frac{1}{2})$ & $(0,\frac{1}{2})$ & $(\frac{1}{2},\frac{1}{2},\frac{1}{2})$ & $(1,0,0)$ & $\frac{3}{2}$ & $-1$ & $-1$ & $0$ \\
        $\B{\lambda}_{\dot{-}}$& $(\frac{3}{2},-\frac{1}{2},\frac{1}{2})$ & $(0,-\frac{1}{2})$ & $(\frac{1}{2},\frac{1}{2},\frac{1}{2})$ & $(1,0,0)$ & $\frac{3}{2}$ & $-1$ & $-1$ & $0$ \\\hline\hline
        $\partial_{+\dot{+}}$ & $(1,1,0)$ & $(\frac{1}{2},\frac{1}{2})$ & $(0,0,0)$ & $(0,0,0)$ & $1$ & $0$ & $0$ & $0$ \\
        $\partial_{+\dot{-}}$ & $(1,0,1)$ & $(\frac{1}{2},-\frac{1}{2})$ & $(0,0,0)$ & $(0,0,0)$ & $1$ & $0$ & $0$ & $0$ \\
    \end{tabular}
    \caption{Charges of the elementary fields in $\C{N}=4$ SYM and $\C{N}=1$ IR SCFT.}
    \label{tab: Listing free BPS letters for LS theory}
\end{table}

On the field-theory side, the superconformal index of $\mathcal{N}=4$ SYM has been extensively studied using various approaches, including the Cardy limit \cite{Choi:2018hmj}, Bethe–Ansatz-type formulas \cite{Benini:2018ywd}, and finite-$N$ computations \cite{Choi:2021rxi, Cabo-Bizet:2019eaf}. These techniques reproduce the corresponding gravity-side results, most notably the black hole entropy in $AdS_5$ \cite{Cabo-Bizet:2018ehj}, and they extend to more general holographic $\C{N}=1$ SCFTs (for instance see \cite{Choi:2023tiq} and references therein). In particular, the large-$N$ saddle-point analysis of the superconformal index of the IR SCFT reveals confining and deconfining phase transitions, with a free energy scaling as $O(N^2)$, dual to the Hawking–Page transitions.

%%%%%%%%%%%%%%%%%%%%%%%%%%%%%%%%%%%%%%%%%
%%%%%%%%%%%%%%%%%%%%%%%%%%%%%%%%%%%%%%%%%
%%%%%%%%%%%%%%%%%%%%%%%%%%%%%%%%%%%%%%%%%
%%%%%%%%%%%%%%%%%%%%%%%%%%%%%%%%%%%%%%%%%
%%%%%%%%%%%%%%%%%%%%%%%%%%%%%%%%%%%%%%%%%
%%%%%%%%%%%%%%%%%%%%%%%%%%%%%%%%%%%%%%%%%
\section{Supercharge cohomology in $\C{N}=1$ SCFT}\label{section 3}
% (\textbf{JC: something like generalization of Romelsberger's prescription: use UV(=free) letters and assign IR R-charges. Here, use UV letters and near-UV Q-actions and assign IR R-charges.})
% (\textbf{JC: conceptual explanations})
% In this section, we define our supercharge cohomology problem. 

Since the $\C{N}=1$ IR SCFT we study in this paper is intrinsically strongly coupled and has no known simple Lagrangian description, we must instead work with the UV Lagrangian.
Accordingly, we define the supercharge cohomology using the classical supercharge of the (mass-deformed) UV Lagrangian. 
% Here, we mean UV Lagrangian by the mass-deformed $\C{N}=4$ SYM Lagrangian discussed in the previous section.
We construct the cochains entirely from the UV free-BPS letters. This is analogous to Romelsberger’s prescription for computing the superconformal index \cite{Romelsberger:2005eg,Romelsberger:2007ec}, which uses UV letters and assigns IR R-charges to them (even though the UV letters might not satisfy the IR BPS bound $\Delta_{IR}-2j_1+\frac{3}{2}r_{IR}=0$).\footnote{
For example, the chiralini of LS SCFT $\psi_{+a}$ has $\Delta_{IR}=\frac{5}{4}$, $j_1=\frac{1}{2}$, and $r_{IR}=-\frac{1}{2}$; therefore, $\Delta-2j_1+\frac{3}{2}r_{IR}=-\frac{1}{2}$. However, they are related to the UV BPS letters \eqref{eqn: UV to IR transf}, and their contribution to the superconformal index is very important.
}

% In our case, we use UV letters to represent IR BPS operators.
% However, we do not assign IR R-charges to the UV letters. 
% Rather, we introduce a \emph{change of variables} (field redefinitions) from UV letters to `\emph{IR letters}'.
% If we treat the mass parameter $M$ as a background chiral superfield, $M$ does not break the UV $U(1)_R$ symmetry, and it can be identified with the IR $U(1)_R$. (\textbf{JC: not true if we think of $M$ as a background superfield.})
% Then, the IR letters that we will define have the desired IR R-charges.
% This may be regarded as a cohomological `explanation' of why Romelsberger's prescription assigns IR R-charges to the UV letters, because the UV letters are transformed into the letters with IR R-charges.

Since we work with the supercharge $Q$ derived from the classical UV Lagrangian, the $Q$-cohomologies can receive quantum corrections.
There are examples whose cohomologies are not protected by quantum corrections.
One example is pure $\C{N}=1$ SYM. The supercharge cohomology problem of a similar kind was studied in \cite{Budzik:2023xbr} for this theory. It was shown explicitly that cohomologies receive perturbative renormalization. 
% It is further expected to suffer from non-perturbative effects, which might eventually lift all the cohomologies except for the chiral ring operators.
Another example is $\C{N}=4$ SYM with gauge group $SO(7)$.
It was found in \cite{Choi:2025bhi} (based on the S-duality analyses of classical cohomologies of $SO/Sp$ $\C{N}=4$ SYM \cite{Gadde:2025yoa,Chang:2025mqp}) that the quantum correction inferred from the generalized Konishi anomaly \cite{Cachazo:2002ry} lifts a pair of (classical) monotone-fortuitous cohomologies.

Bearing this in mind, we aim to `explain' the superconformal index by finding the corresponding cohomologies.
% we consider the classical cohomology problem \textit{near} UV.
If there are more cohomologies than the index requires, we keep in mind that some of the cohomologies might pair up and be lifted.
Note that the pairing can be monotone-fortuitous. This means that the graviton index can also receive quantum corrections.

%%%%%%%%%%%%%%%%%%%%%%%%%%%%%%%%%%%%%%%%%
%%%%%%%%%%%%%%%%%%%%%%%%%%%%%%%%%%%%%%%%%
%%%%%%%%%%%%%%%%%%%%%%%%%%%%%%%%%%%%%%%%%
%%%%%%%%%%%%%%%%%%%%%%%%%%%%%%%%%%%%%%%%%
\subsection{$Q$-action and change of variables} \label{sub:change of variables}
We begin with the classical UV definition of the $Q$-action. 
From now on, for the ease of notation, we denote $(\bar{\phi}^m, \psi_{a+}, F_{++}, \B{\lambda}_{\D{\alpha}}, D_{+\D{\alpha}})$ by $(\phi^a, \psi_a, f, \lambda_\alpha, D_\alpha)$.
One can always unambiguously recover the original notation since our supercharge cohomology problem never uses other (non BPS) fields. 

The $Q$-action of mass-deformed $\mathcal{N}=4$ SYM in the UV is given by
\begin{equation}\label{eq: UV Q}
    \begin{aligned}
        [Q,\phi^m]&=0\\
        \{Q,\psi_{m}\}&=\frac{1}{2}\epsilon_{mnp}[\phi^n,\phi^p]+M\delta_{m,3}\phi^3\\
        [Q,f]&=[\phi^m,\psi_{m}]\\
        \{Q,\lambda_{\alpha}\}&=0\\
        [Q,D_{\alpha}]\bullet&=[\lambda_{\alpha},\bullet\}
    \end{aligned}
\end{equation}
Let us define $(X,Y,Z)\equiv ({\phi}^1,{\phi}^2,{\phi}^3)$.
Then,
\begin{align}
    Z=\frac{1}{M}(Q\psi_{3}-[X,Y])\;,
\end{align}
Substituting into $\{Q,\psi_{a}\}$ for $a=1,2$ gives
\begin{equation}
    \Big\{Q,\Big(\psi_{a}+\frac{1}{M}[{\phi}_a,\psi_{3}]\Big)\Big\}=-\frac{1}{M}[{\phi}_a,{\phi}^b{\phi}_b]
\end{equation}
In the above manipulation, we have raised and lowered $SU(2)_F$ flavor indices with $\epsilon_{ab}$ and $\epsilon^{ab}$ where $\epsilon^{12}=-\epsilon_{12}=1$. For example, a flavor singlet is obtained by contracting indices as follows:
\begin{equation}
    \epsilon_{ab}\phi^a\phi^b=\phi^a\phi_a\equiv(\phi\phi),\qquad \epsilon_{ab}\psi^a\psi^b=\psi^a\psi_a\equiv(\psi\psi)
\end{equation}
In particular, $\epsilon^{12}{\phi}_2={\phi}^1={\phi}_2$ and $\epsilon^{21}{\phi}_1={\phi}^2=-{\phi}_1$.
 
Similarly, for $[Q,f]$, we find
\begin{equation}
    \Big[Q,f-\frac{1}{M}\psi_{3}\psi_{3}\Big]=\Big[{\phi}^a,\psi_{a}+\frac{1}{M}[{\phi}_a,\psi_{3}]\Big]
\end{equation}
We now define
\begin{equation}\label{eqn: UV to IR transf}
    \begin{split}
        {\phi}^a_{IR}&\equiv 
        M^{-1/4}{\phi}^a_{UV}\ ,\\
        \psi_{IR\,a}&\equiv 2M^{1/4}\Big(\psi_{UV\,a}+\frac{1}{M}[{\phi}_{UV\,a},\psi_{3}]\Big)\ ,\\
        f_{IR}&\equiv -2\Big(f_{UV}-\frac{1}{M}\psi_{3}\psi_{3}\Big)\ ,
    \end{split}
\end{equation}
and $\lambda_{IR}\equiv\lambda_{UV}$, $D_{IR}\equiv D_{UV}$.
The subscript $IR$ is a \emph{mnemonic device} rather than truly denoting the IR quantum fields.
As we mentioned before, we are still working \emph{near} UV, but we will see in the next subsection that this change of variables `integrates out' the massive degrees of freedom in a cohomological sense.
Then, we obtain
\begin{equation}\label{eqn: IR Q-action}
    \begin{aligned}
        [Q,{\phi}_{IR}^a]&=0\\
        \{Q,\psi_{IR\, a}\}&=-2[{\phi}_{IR\,a},{\phi}_{IR}^b{\phi}_{IR\,b}]\\
        [Q,f_{IR}]&=-[{\phi}^a_{IR},\psi_{IR\,a}]\\
        \{Q,{\lambda}_{IR\,\alpha}\}&=0\\
        [Q,D_{IR\,\alpha}]\bullet&=[\lambda_{IR\,\alpha},\bullet\}
    \end{aligned}
\end{equation}
This is the $Q$-action that will be used throughout this paper for the IR SCFT.
Although the real $Q$-action on the IR quantum fields would be much more complicated,\footnote{
$Q_-\psi_{+a}\propto F_a$ and $Q_-F_{++}\propto D_{+\dot{\alpha}}\bar{\lambda}^{\dot{\alpha}}$ are still valid, where $F_a$ is the auxiliary field in the chiral multiplet $\Phi_a$, because these are off-shell relations. However, the on-shell relations are modified due to the non-trivial K\"ahler potential. Schematically, 
\begin{align*}
    \epsilon^{ab}F_b=
    g^{a\bar{b}}\partial_{\bar{b}}\B{W}+
    \Gamma^{a}_{bc}(\psi^{\alpha b}\psi^c_{\alpha})
    \;,\;\;\;
    D_{\alpha\dot{\alpha}}\bar{\lambda}^{\dot{\alpha}}
    =[\B{\Phi}^a,g_{b\bar{a}}\psi_{\alpha}^b]\;,
\end{align*}
where $g_{a\bar{b}}\equiv \frac{\partial^2 K}{\partial\Phi^a\partial\B{\Phi}^b}=\epsilon_{ac}\frac{\partial^2 K}{\partial\Phi_c\partial\B{\Phi}^b}$ and $\Gamma^a_{bc}=g^{a\bar{d}}\partial_{b}g_{c\bar{d}}$.
Here, the index contraction includes the (adjoint) gauge index contractions, except for the index $\bar{a}$ in the commutator, in which $(g_{b\bar{a}}\psi^{b}_{\alpha})^i_j=\frac{\partial^2 K}{\partial(\B{\Phi}^a)^i_j\partial(\Phi^b)^k_l}(\psi^b_\alpha)^k_l$.
Besides these, there can be quantum corrections due to the renormalization of composite operators.
} we will see that this near UV cohomology still satisfies an abundance of desirable properties. 
For example, we can find the single-trace graviton cohomologies corresponding to the KK-particles in the Pilch-Warner background.
Also, this $Q$-cohomology `explains' the superconformal index of the IR SCFT. Namely, the number of bosonic cohomologies minus the number of fermionic cohomologies reproduces the index (which should be the case to be consistent with Romelsberger's prescription).
% (\textbf{JC: graviton spectrum, integrates out massive fields, explains the superconformal index—always been able to find a representative})

We normalized ${\phi}_{IR}$ so that the coefficient of the IR quartic superpotential is of order one $\sim M^0$. 
% Since scaling dimensions do not add up in the IR while R-charges remain additive, one might view that $M$ has R-charge $-\frac{2}{3}$. 
It is convenient to assign R-charge to $M$ as $-\frac{2}{3}$, viewing $M$ as a background anti-chiral superfield.
Then UV $U(1)_R$ is (spuriously) unbroken and differs from the IR $U(1)_R$ by a (spurious) flavor $U(1)_S$.
The abelian charges of UV and IR letters are listed in Table \ref{tab: spurious charges}.
The UV R-charge is related to the IR R-charge as $r^{UV}=r^{IR}-\frac{S}{6}$.
\begin{table}[H]
    \centering
    \begin{tabular}{c||c|c|c|c|c||c|c|c|c}
      & $\phi^a$ & $\psi_a$ & $Z$ & $\psi_3$ & $M$ & $\phi^a_{IR}$ & $\psi_{IR\,a}$ & $\zeta$ & $\chi$ \\ \hline\hline
      $r^{UV}$ & $-\frac{2}{3}$ & $-\frac{1}{3}$ & $-\frac{2}{3}$ & $-\frac{1}{3}$ & $-\frac{2}{3}$ & $-\frac{1}{2}$ & $-\frac{1}{2}$ & $-1$ & $0$ \\ \hline
      $r^{IR}$ & $-\frac{1}{2}$ & $-\frac{1}{2}$ & $-1$ & $0$ & $0$ & $-\frac{1}{2}$ & $-\frac{1}{2}$ & $-1$ & $0$ \\ \hline
      $S$ & $1$ & $-1$ & $-2$ & $2$ & $4$ & $0$ & $0$ & $0$ & $0$ \\ \hline
    \end{tabular}
    \caption{Spurious charges $(r^{UV},S)$ and the IR superconformal R-charge $r^{IR}$. The IR heavy fields $(\zeta,\chi)$ are defined in equation \eqref{eq: heavy fields}.}
    \label{tab: spurious charges}
\end{table}\vspace{-0.3cm}
\noindent Note that the IR letters (including $\lambda_\alpha,D_\alpha,f_{IR}$) are all neutral under $U(1)_S$, and hence $r^{UV}=r^{IR}$ for them.
Therefore, working with IR coordinates enables us to identify UV and IR R-symmetries, with the understanding that $M$ carries the UV R-charge.
% Indeed,
% \begin{equation*}
%     R[{\phi}_{IR}]=-\frac{1}{2}=-\frac{1}{4}\left(-\frac{2}{3}\right)-\frac{2}{3}=-\frac{1}{4}R[M]+R[{\phi}_{UV}] \ .
% \end{equation*}
% Similarly, since $R[\psi_{IR}]=\frac{1}{2}-1=-\frac{1}{2}$ and $R[\psi_{UV}]=\frac{2}{3}-1=-\frac{1}{3}$, a factor of $M^{1/4}$ appears in front of $\psi_{UV}$. 

% Assigning an $R$-charge to $M$ is reminiscent of spurion analysis, where the mass $M$ is treated as a scalar field at very high energy scales, though this viewpoint will not play a significant role in our discussion.

% (\textbf{JC: assigning R-charge to $M$ looks like viewing $M$ as a chiral superfield. But we are not doing that e.g. there is no $\psi_{M\alpha}$. Conceptually slightly confusing.})

The $Q$-action \eqref{eqn: IR Q-action} could alternatively be derived from another UV gauge theory, whose matter content consists of two adjoint chiral multiplets with the standard quadratic K\"ahler potential and the quartic superpotential $W\propto \Tr [\Phi_1,\Phi_2]^2$. 
Due to the quartic superpotential, this theory is not renormalizable. However, it is a sensible theory as a Wilsonian effective field theory that gives rise to the desired IR CFT.
More precisely, this alternative UV theory flows to the same conformal manifold, presumably with a different coefficient in front of the superpotential. 
% In this case, the superpotential is irrelevant, and hence the coefficient in front of it becomes infinitesimal in the IR. In the mass-deformed $\C{N}=4$ SYM case, the coefficient is hard to predict, but 
% $K\propto\Tr \B{\Phi}^a e^{[V,\cdot]}\Phi_a$, where $\B{\Phi}^a=(\Phi_a)^{\dag}$.

% \paragraph{From (near) UV to IR}
%%%%%%%%%%%%%%%%%%%%%%%%%%%%%%%%%%%%%%%%%
%%%%%%%%%%%%%%%%%%%%%%%%%%%%%%%%%%%%%%%%%
%%%%%%%%%%%%%%%%%%%%%%%%%%%%%%%%%%%%%%%%%
%%%%%%%%%%%%%%%%%%%%%%%%%%%%%%%%%%%%%%%%%
\subsection{From UV to IR} \label{sub: integrating out}
Going from (mnemonic) IR to (near) UV is straightforward: substitute \eqref{eqn: UV to IR transf} into an IR operator. As we will see later, the nontrivial IR cohomologies become nontrivial UV cohomologies (with small mass), while trivial cohomologies map to trivial ones.

However, the other way around is nontrivial. This is because one has to consistently turn off the UV fields to arrive at the IR expression. This is a non-linear truncation that proceeds as follows. One first re-expresses the (mass deformed) UV operators, originally written in the ``UV coordinate'' $({\phi}^m,\psi_m,f)$, in terms of the ``IR coordinate'' $({\phi}_{IR}^a,\psi_{IR\,a},f_{IR}, \zeta, \chi)$. Here, we defined
\begin{align}\label{eq: heavy fields}
    \zeta\equiv M^{1/2}\left( Z+\frac{[X,Y]}{M} \right),\qquad
    \chi\equiv M^{-1/2}\psi_3 \ .
\end{align}
Since $M\to t^2$, both $\zeta$ and $\chi$ scale as $t^3$, canceling each other in the index.
The letters $\lambda_\alpha$ and $D_\alpha$ are omitted, since the transformation acts on them trivially.
This \textit{change of coordinates} is always possible, and one can simply substitute the following expressions:
\begin{equation}
\begin{aligned}
    &\phi^a_{UV}=M^{1/4}\phi^a_{IR}\;,\;\;
    \psi_{UV\,a}=M^{-1/4}\Big(\frac{1}{2}\psi_{IR\,a}-[\phi_{IR\,a},\chi]\Big)\;,\\
    f_{UV}&=-\frac{1}{2}f_{IR}+\chi^2\;,\;\;
    Z=M^{-1/2}\big(\zeta-[\phi^1_{IR},\phi^2_{IR}]\big)\;,\;\;
    \psi_3=M^{1/2}\chi\;.
\end{aligned}
\end{equation}

This choice of IR coordinates has a useful property: the $Q$-action becomes block-diagonal,
\begin{equation}
\begin{aligned}
    Q\psi_{IR\,a}
    =-2[\phi_{IR\,a},\phi^b_{IR}\phi_{IR\,b}],\quad
    &Qf_{IR}
    =-[\phi^a_{IR},\psi_{IR\,a}],\\
    [Q,D_\alpha]\bullet=[\lambda_\alpha,\bullet\},\quad
    &Q\chi= \zeta \ .
\end{aligned}
\end{equation}
Thus the operators split into two sectors: those involving $\zeta$, $\chi$ and their derivatives, which we call heavy, and those built only from $({\phi}^a_{IR},\psi_{IR,a},f_{IR},D_{\alpha},\lambda_\alpha)$, which we call light. Schematically, the $Q$-action acts as
% Namely, it is of the following form. If we call operators containing the letters $\zeta, \chi$ and their derivatives `heavy', and operators made only of $({\phi}^a_{IR},\psi_{IR\,a}, f_{IR},D_{\alpha},\lambda_\alpha)$ `light', the $Q$ action is schematically
\begin{align}
    Q(\text{heavy})=(\text{heavy}),\quad
    Q(\text{light})=(\text{light}) \ .
\end{align}
Just as in the BMN truncation (which will be explained in the subsection \ref{sub:BMN}), one can therefore set the heavy fields, $\zeta,\chi$ and their derivatives, to zero:
% one can therefore consistently turn off the heavy fields ($\zeta,\chi$ and their derivatives):
\begin{align}
    \zeta=0,\quad
    \chi=0 \ .
\end{align} 
This corresponds to the IR setting.
% This is, of course, the IR setting.

Furthermore, one can show that every $Q$-cohomology class \emph{always} admits a representative without heavy letters.
This can be shown by introducing an operator $S_H$, which acts only on $\zeta$ and its derivatives as
\begin{align}
    S_H\zeta = \chi\;,\;\;
    S_H\partial_\alpha\zeta = \partial_\alpha\chi
    \;,\;\;
    S_H\partial_\alpha\partial_\beta\zeta 
    = \partial_\alpha\partial_\beta\chi\;,\;...
\end{align}
As a differential operator, $S_H$ can be written as
\begin{align}
    S_H=\chi\frac{\partial}{\partial\zeta}
    +\partial_\alpha\chi \frac{\partial}{\partial(\partial_\alpha\zeta)}
    +\partial_\alpha\partial_\beta\chi \frac{\partial}{\partial(\partial_\alpha\partial_\beta\zeta)}
    +\cdots .
\end{align}
The anti-commutator with $Q$ then becomes
\begin{align}
    \{Q,S_H\}=N_H\;,
\end{align}
where
\begin{align}
    N_H=\zeta\frac{\partial}{\partial\zeta}
    +\chi\frac{\partial}{\partial\chi}
    +\partial_\alpha\zeta \frac{\partial}{\partial(\partial_\alpha\zeta)}
    +\partial_\alpha\chi \frac{\partial}{\partial(\partial_\alpha\chi)}
    +\partial_\alpha\partial_\beta\zeta \frac{\partial}{\partial(\partial_\alpha\partial_\beta\zeta)}
    +\partial_\alpha\partial_\beta\chi \frac{\partial}{\partial(\partial_\alpha\partial_\beta\chi)}
    +\cdots 
\end{align}
is the number operator that counts heavy letters. For example, 
\begin{equation}
    N_H \Tr(\chi F_{IR} D_\alpha\zeta)=2\Tr(\chi F_{IR} D_\alpha\zeta) \ .
\end{equation}
Given an operator $O$, one may expand it in eigenstates of $N_H$:
\begin{equation}
    O=O_0+\sum_{k=1}^{n}O_k\ ,
\end{equation}
where $N_H O_k=k O_k$. One can rewrite this as
\begin{equation}
    O=O_0+\{Q,S_H\}O_H\ ,
\end{equation}
where $O_H=\sum_{k=1}^n \frac{1}{k}O_k$. If $O$ is $Q$-closed, then each $O_k$ is $Q$-closed since $[Q,N_H]=0$. In particular, $O_H$ is $Q$-closed, and thus
\begin{align}
    O=O_0+Q(S_H O_H)\ .
\end{align}
In other words, any cohomology class has a representative $O_0$, built solely from light letters.
This may be viewed as the massive fields being `integrated out' in a cohomological sense.

\subsection{Coulomb-type cohomologies and moduli space}\label{sub:coulomb noncoulomb}
% Following earlier works on graviton and black hole cohomologies in $\mathcal{N}=4$ SYM \cite{Choi:2022caq, Choi:2023znd, Choi:2023vdm}, we apply the eigenvalue counting method, which at finite $N$ efficiently identifies graviton cohomologies where trace relations among operators become important. This allows us to isolate non-graviton operators, those that are not products of gravitons, and which exhibit black-hole-like entropy growth.

Now we have a $Q$-action that yields only commutators \eqref{eqn: IR Q-action} because the pair of `heavy' fields $\zeta,\chi$ ($Q\chi=\zeta$) has been integrated out. 
Therefore, it is convenient to define \emph{Coulomb-type} cohomology. We define it as a cohomology class that is non-vanishing when the (adjoint) fields are in the Cartan subalgebra.\footnote{
For other theories where the matter fields are not necessarily in the adjoint representation, one can generalize the definition as cohomologies that are non-vanishing for field configurations that make the classical $Q$-action vanish.
}
A non-Coulomb cohomology is a cohomology class that vanishes when the fields are in the Cartan.
Since the $Q$-action \eqref{eqn: IR Q-action} yields commutators, $Q$-exact operators are always non-Coulomb (as operators), making these definitions (as cohomologies) well-defined.

As mentioned in the introduction, in $\C{N}=4$ SYM with gauge group $SU(N)$, all the known (multi)graviton cohomologies are Coulomb, and all the known non-graviton cohomologies are non-Coulomb.

% (\textbf{JC: Two classification criteria graviton/non-graviton or eigenvalue/non-eigenvalue. Ben diagram? Classification of eigen-noneigen is based on the specific form of the $Q$-aciton. In $\C{N}=4$ SYM, eigenvalue and graviton might be the same thing and non-gravtion=non-eigenvalue. But in LS theory, there exist overlaps between graviton and non-eigenvalue, and between non-gravtion and eigenvalue.})

% In $\mathcal{N}=4$ SYM, graviton primaries in the chiral ring generate entire towers of graviton cohomologies (super-descendants) by acting with the supercharges in the $\mathfrak{psu}(1,2|3)$ commuting subalgebra. The primaries of these multiplets, denoted $S_n$ in \cite{Choi:2023znd, Choi:2023vdm}, are built from commuting scalar operators $\phi^m$, whose gauge-invariant data are fully captured by their eigenvalues. Since the supercharges act linearly, the descendants likewise remain diagonal as well. Thus, graviton cohomologies are completely characterized by the eigenvalues of the matrix-valued fields. In this sense, ``graviton'' and ``eigenvalue'' can be regarded as interchangeable notions, while ``non-graviton'' corresponds to ``non-eigenvalue''.

In the LS SCFT, the identifications of Coulomb with graviton and non-Coulomb with non-graviton are hopelessly broken.
We encounter four distinct types of cohomologies:
\begin{itemize}
    \item Coulomb graviton: $u_n,v_n,w_0$ 
    \item Coulomb non-graviton: $O_7,O_9,O_{12},O_{14},\cdots +(\mathrm{hairs})$ (in $SU(2)$)
    \item non-Coulomb graviton: $w_0^2$ (in $SU(2)$), $w_{1\leq n\leq N-2}$ 
    \item non-Coulomb non-graviton: $O_{16}$ (in $SU(2)$)
\end{itemize}
For explicit expressions of the examples listed above, see \eqref{eqn: Single graviton in BMN}, Section \ref{sub: Coulomb nongrav}, and Section~\ref{sub:nonCoulombnongraviton}.

At this point, it would be remiss not to discuss how the Coulomb cohomologies are related to the moduli space of vacua.
The (IR) scalar gravitons $u_n$ are written as follows.
\begin{align}
    u_n^{a_1\cdots a_{n+2}}=\Tr \phi^{(a_1}\cdots \phi^{a_{n+2})}\;,
\end{align}
Here we omit the $IR$ subscripts, and we will continue to do so until Section \ref{section:IR cohomology in UV coordinates}, as far as the distinction between the IR and UV expressions is clear from the context.
These operators are in the (anti) chiral ring, and all the chiral ring operators made of scalars are multi-traces of $u_n$.
This is because whenever there is an antisymmetric pair of scalars in the trace, it becomes $Q$-exact.
Unlike in the $\C{N}=4$ case, this fact is slightly nontrivial because the $Q$-exact combinations are double commutators $Q\psi_1\sim [Y,[X,Y]]$, $Q\psi_2\sim [X,[X,Y]]$ instead of single commutators.

One can show this as follows. 
It suffices to show that the single trace cohomologies $O^{a_1\cdots a_{n+2}}=\Tr \phi^{a_1}\cdots \phi^{a_{n+2}}$ can always be represented as $u_n$.
Suppose that $O^{a_1\cdots a_{n+2}}$ is in spin $n/2$ representation of $SU(2)$ flavor.
This means that there is one pair of antisymmetrized scalars and others are all symmetrized.
Such operator is a linear combination of $\Tr \phi^{(a_1}\cdots\phi^{a_n)}[X,Y]$, which is identically vanishing.
The vanishing can be verified by taking $a_1=...=a_n=1$. Then, $\Tr X^n[X,Y]=\Tr Y[X^n,X]=0$.
Other components of the $n/2$ representation obey the same identity because $[X,Y]$ is a singlet.
Therefore, when there is one antisymmetric pair and others all symmetric, it is identically vanishing.
As a next step, consider the spin $(n-2)/2$ representation of $O^{a_1\cdots a_{n+2}}$, which has two pairs of antisymmetrized scalars.
It is a linear combination of $\Tr \phi^{(a_1}\cdots \phi^{a_k}[X,Y]\phi^{a_{k+1}}\cdots \phi^{a_{n-2})}[X,Y]$.
Since $\phi^a$ and $[X,Y]$ commute up to $Q$-exactness, we focus on $\Tr \phi^{(a_1}\cdots \phi^{a_{n-2})}[X,Y]^2$.
Setting $a_1=...=a_{n-2}=1$, we see that these are $Q$-exact:
$$\Tr X^{m}[X,Y]^2 =\Tr Y [X^m[X,Y],X]=-\Tr YX^m[X,[X,Y]]\sim Q\Tr Y X^m \psi_2\;.$$
This identity can be shown for the whole representation of spin $(n-2)/2$ of flavor $SU(2)$.
Therefore, when there are two pairs of antisymmetric scalars, it becomes $Q$-exact.
Similarly, when there are three pairs of antisymmetric scalars, one can focus on terms like $\Tr\phi^{(a_1}\cdots \phi^{a_{n-4})}[X,Y]^3$ and this is $Q$-exact for the similar reason:
$$\Tr X^m[X,Y]^3=\Tr Y[X^m[X,Y]^2,X]=-\Tr YX^m[X,[X,Y]^2]\sim Q\Tr Y X^m\{[X,Y],\psi_2\}.$$
This can be repeated for any smaller representations, proving the claim.

% (\textbf{JC: chiral primaries being described by eigenvalues is nontrivial because $Q\psi_a$'s are double commutators. Maybe better to explain how this is possible. Also, examples like $\Tr([X,Y]^2)=\Tr(X[Y,[X,Y]])\propto Q\Tr(X\psi_1)$ might be nice to display.})

% At this point, it is helpful to view chiral ring primary operators as objects whose vacuum expectation values parametrize the moduli space of the gauge theory. In $SU(N)$ $\C{N}=4$ SYM, the $F$-term and $D$-term conditions guarantee that the three complex adjoint scalars can be simultaneously diagonalized. The resulting `classical' Coulomb branch moduli space (far away from the origin) is
% \begin{equation}
%     \C{M}_{UV}=\mathbb{R}^{6(N-1)}/S_N
% \end{equation}
% The gauge-invariant chiral ring operators, such as $\Tr(\Phi_1^k)$, are then described by symmetric polynomials of the eigenvalues of the adjoint scalars. Antisymmetric combinations vanish in cohomology, since they are $Q$-exact: $[\bar{\phi}^m,\bar{\phi}^n]\sim \epsilon^{mnp}Q\psi_{p+}$.

Since the vacuum expectation values (VEVs) of these (anti) chiral ring operators parametrize the moduli space of vacua, the fact that the scalar fields are always totally symmetrized in the trace roughly implies that the moduli space of vacua is described by the eigenvalues of the two scalar fields.
Namely, we expect the Coulomb branch moduli space to be
\begin{equation}
    V_4^{(N-1)}/S_N\;,
\end{equation}
where $V_4$ is the space in which each eigenvalues of $(X,Y)$ live in, and is topologically $\mathbb{R}^4$. Here, $S_N$ act on the eigenvalues of $(X,Y)$ simultaneously. 
However, it is difficult to verify this, as we do not know the Lagrangian of the IR SCFT, let alone the quantum corrections to the moduli space.
% In this case, it is no longer obvious that the chiral primaries can be described in terms of eigenvalues, since the nontrivial $F$-term (and $D$-term) conditions impose relations between them. This feature is reflected in the double-commutator structure of the $Q$-action in the IR theory, as shown in \eqref{eqn: IR Q-action}. However, still, it may hold at least cohomologically, because for example $\Tr([X,Y]^2)=\Tr(X[Y,[X,Y]])\sim Q\Tr(X\psi_1)$, which is again $Q$-exact.

One method would be to conduct a probe D3-brane analysis in the Pilch-Warner background \cite{Pilch:2000ej}, which can be interpreted as pulling a single D3-brane away from the $N$ stack of them, breaking the gauge group from $SU(N)\to SU(N-1)\times U(1)$.
In \cite{Johnson:2000ic}, a probe analysis was done in another type $\rm IIB$ supergravity solution found by Pilch and Warner \cite{Pilch:2000fu}, that describes the RG flow, interpolating the UV and IR fixed points.
The motion of this probe in the transverse directions explores the moduli space of vacua.
In the UV limit, where the internal metric reduces to that of a round $S^5$ and there is no warping factor in front of $AdS_5$, the D3-brane potential vanishes and the probed moduli space is flat $\mathbb{R}^6$. In the IR, due to the warping factor in front of the $AdS_5$ metric, two directions get lifted, and the moduli space metric probed by a D3-brane takes the following form:  
\begin{align}
    ds^2_{\mathcal{M}_{IR}}=\frac{3}{4}du^2+u^2
    \Big(\sigma_1^2+\sigma_2^2+\frac{4}{3}\sigma_3^2\Big)\;,
\end{align}
in some coordinates \cite{Johnson:2000ic}. $\sigma_i$ are the $SU(2)$ left-invariant one-forms satisfying $d\sigma_i =\epsilon_{ijk}\sigma_j\wedge\sigma_k$. This is in agreement with the field theory expectation: it is four dimensional, topologically $\mathbb{R}^{4}$, and preserves $SU(2)\times U(1)_R$.
Furthermore, it has a conical singularity at the origin because of the factor $\frac{4}{3}$.

Regarding the existence of non-Coulomb gravitons, we speculate that it is closely related to the conical singularity (besides the usual orbifold singularity) at the origin of the moduli space.
\subsection{BMN truncation}\label{sub:BMN}
We focus on a specific subsector of the cohomology problem, known as the BMN sector~\cite{Berenstein:2002jq,Kim:2003rza,Choi:2023znd} (also referred to as the no-derivative sector \cite{Budzik:2023vtr}). In the $\C{N}=4$ SYM case, this sector consists of operators constructed from the following letters:
\begin{equation}\label{eqn: BMN letters}
    {\phi}^m,\qquad \psi_{m},\qquad f, \qquad m=1,2,3\ .
\end{equation}
$i.e.$ no derivatives and gaugino. This sector is closed under the 1-loop $Q$-action:
\begin{equation}\label{eqn: Q-action in BMN}
    [Q,\phi^m]=0,\qquad \{Q,\psi_{m}\}=\frac{1}{2}\epsilon_{mnp}[\phi^m,\phi^n],\qquad [Q,f]=[\phi^m,\psi_{m}]\ ,
\end{equation}
A sector is said to be closed if no operator is mapped out of the sector by the $Q$-action, and no operator from outside the sector is mapped into it.
In this sense, the BMN truncation is consistent with the 1-loop $Q$-cohomology problem. This allows one to compute the index within the BMN sector (which is valid at least for the 1-loop BPS spectrum) by manually discarding non-BMN letters. Although this truncation removes infinitely many letters ($i.e.$ derivatives), it has been shown that the entropy of the BMN sector still scales as $\mathcal{O}(N^2)$ in the large-$N$ limit \cite{Choi:2023vdm}.

In the BMN sector, the single-letter index reduces to
\begin{equation}
    i_s^{UV}=1-\left(1-t^2v\right)\left(1-\frac{t^2w}{v}\right)\left(1-\frac{t^2}{w}\right)=1-(1-x)(1-y)(1-z)
\end{equation}
For simplicity, let us set $v = w = 1$ (equivalently, $x = y = z = t^2$), so that
\begin{equation}
    i_s^{UV}=3t^2-3t^4+t^6\ ,
\end{equation}
where the terms correspond to the contributions from three scalars, three fermions, and one field strength in the BMN sector. This one-parameter unrefinement further reduces the index to
\begin{equation}
    \C{I}^{UV}=\Tr(-1)^F t^{3(2j-r^{UV})}=\Tr(-1)^F t^{6(j+R)},\qquad R\equiv \frac{1}{3}(R_1+R_2+R_3)\ ,
\end{equation}
where $6(R+j)$ is quantized to be an integer.

To be explicit, we compute the $SU(2)$ BMN index using the matrix integral expression:
\begin{align}
    \C{I}^{SU(2)}&=\frac{1}{2!}\int_0^{2\pi}\prod_{a=1}^2\frac{d\alpha_a}{2\pi }\exp\bigg[\sum_{a\neq b}\sum_{n=1}^\infty\frac{i_s(t^n)-1}{n}e^{in\alpha_{ab}}\bigg]\exp\bigg[\sum_{n=1}^\infty\frac{i_s(t^n)}{n}\bigg]\ .
\end{align}
where $\alpha_{ab}=\alpha_a-\alpha_b$. After evaluating the contour integral by summing over all residues, we obtain
\begin{equation}
    \begin{split}
        \C{I}^{UV,\,SU(2)}_{\text{BMN}}(t)&=\Big[1+3t^2+12t^4+20t^6+42t^8+75t^{12}+66t^{14}+81t^{16}+55t^{18}\\
        &\qquad+54t^{20}+27t^{22}+19t^{24}+6t^{26}+3t^{28}\Big]\frac{(1-t^2)^3}{(1-t^8)^3(1-t^{12})}\ .
    \end{split}
\end{equation}
Both black hole and graviton cohomologies made with the letters \eqref{eqn: BMN letters} contribute to this index.

Let us next consider the IR SCFT, where the BMN truncation requires using only the following letters:
\begin{equation}\label{eqn: LS Q-action}
    \phi^a,\qquad \psi_{a},\qquad f,\qquad a=1,2 \ .
\end{equation}
In the IR, the $Q$-action
\begin{equation}\label{eqn: Q-action of BMN in LS theory}
    [Q,\phi^a]=0,\quad \{Q,\psi_{a}\}=-2[\phi_a,\phi^b\phi_b],\quad [Q,f]=-[\phi^a,\psi_{a}]\ .
\end{equation}
form a closed set under the action of $Q$. The single-letter index is now given by
\begin{equation}
    i_s^{IR}=1-\left(1-\frac{t^{3/2}}{s}\right)\left(1-st^{3/2}\right)\left(1-t^3\right) \ .
\end{equation}
Turning off the $SU(2)$ flavor symmetry for simplicity yields
\begin{equation}
    i_s^{IR}=2t^{3/2}-2t^{9/2}+t^6\ .
\end{equation}
In this case, the index is given by
\begin{equation}
    \C{I}^{IR}=\Tr(-1)^F t^{3(2j-r^{IR})}=\Tr(-1)^F t^{6(j+R^{IR})},\qquad R^{IR}\equiv \frac{1}{4}(R_1+R_2+2R_3) \ .
\end{equation}
Similarly, we compute
\begin{equation}
    \begin{split}
        \C{I}^{IR,\,SU(2)}_{\text{BMN}}&=\Big[1+3t^3+2t^{9/2}+2t^6+4t^{15/2}
        +3t^9+2t^{21/2}+4t^{12}+2t^{27/2}+2t^{33/2}\Big]
        \\
        &\quad\times
        \frac{(1-t^{9/2})^2}{(1-t^{15/2})^2(1-t^{12})}\ .
    \end{split}
\end{equation}

%%%%%%%%%%%%%%%%%%%%%%%%%%%%%%%%%%%%%%%%%
%%%%%%%%%%%%%%%%%%%%%%%%%%%%%%%%%%%%%%%%%
%%%%%%%%%%%%%%%%%%%%%%%%%%%%%%%%%%%%%%%%%
%%%%%%%%%%%%%%%%%%%%%%%%%%%%%%%%%%%%%%%%%
%%%%%%%%%%%%%%%%%%%%%%%%%%%%%%%%%%%%%%%%%
%%%%%%%%%%%%%%%%%%%%%%%%%%%%%%%%%%%%%%%%%
\section{Black hole cohomology in $\C{N}=1$ SCFT}\label{section 4}
We now turn to the finite $N$ black hole cohomology problem in the IR SCFT. The first step is to identify all the graviton-type operators. 

%%%%%%%%%%%%%%%%%%%%%%%%%%%%%%%%%%%%%%%%%
%%%%%%%%%%%%%%%%%%%%%%%%%%%%%%%%%%%%%%%%%
%%%%%%%%%%%%%%%%%%%%%%%%%%%%%%%%%%%%%%%%%
%%%%%%%%%%%%%%%%%%%%%%%%%%%%%%%%%%%%%%%%%
\subsection{Gravitons in IR SCFT}\label{sub:gravitons}
% In $\mathcal{N}=4$ SYM, single-trace graviton operators are identified with chiral primary operators (anti-chiral part of highest weight of $B_1\bar{B}_1[0;0]_{\Delta=n}^{(0,n,0)}$) and their superconformal descendants. Upon choosing a supercharge $Q$, the subalgebra of $\mathfrak{psu}(2,2|4)$ that (anti-)commutes with $Q$ is $\mathfrak{psu}(1,2|3)$. The $1/16$-BPS graviton operators are then generated by acting with the supercharges in $\mathfrak{psu}(1,2|3)$ and BPS-derivatives on the chiral primaries. These graviton-type cohomologies are in one-to-one correspondence with the Kaluza-Klein BPS-particles of type IIB supergravity on $AdS_5 \times S^5$ \cite{Gunaydin:1984fk, Kim:1985ez}.
% In $\mathcal{N}=4$ SYM, single trace graviton operators are in the $B_1\bar{B}_1[0;0]_{\Delta=n}^{(0,n,0)}$ representation of $PSU(2,2|4)$. Here, we follow the convention of \cite{Cordova:2016emh}. 

In $\mathcal{N}=1$ superconformal theories, there is no supercharge of $\mathfrak{su}(2,2|1)$ that anti-commutes with the chosen supercharge $Q = Q_-$. As a result, unlike the $\C{N}=4$ case, graviton spectrum consists of several different superconformal multiplets, and identifying (primary) single trace graviton operators beyond the (anti-)chiral ring is nontrivial.\footnote{
In $\mathcal{N}=4$ SYM, all the single trace graviton operators are in the $B_1\bar{B}_1[0;0]_{\Delta=n}^{(0,n,0)}$ representation of $PSU(2,2|4)$, and therefore it is relatively easy to construct them.
$1/16$-BPS part of them with respect to $Q=Q^4_-$ is obtained as follows. Among the highest weight of $B_1\bar{B}_1[0;0]^{(0,n,0)}_n$, anti-chiral part is given as $\Tr \phi^{(m_1}\cdots \phi^{m_n)}$. By acting generators of (anti-)commuting subgroup $PSU(1,2|3)$, which consists of 9 supercharges and two holomorphic derivatives, one obtains the whole $1/16$-BPS single trace gravitons.
} 

Based on the Kaluza-Klein (KK) spectrum obtained in \cite{Bobev:2020lsk}, we found explicit representatives of the single-trace graviton cohomologies. The KK spectrum can be decomposed into irreducible superconformal multiplets, and we list part of shortened multiplets that are relevant for our cohomology in Table~\ref{tab: Single-trace graviton} (irrelevant ones are their complex conjugate multiplets).
We also list the corresponding cohomology representative in each multiplet.

Let us briefly explain the representations of the 4d $\C{N}=1$ superconformal algebra $\mathfrak{su}(2,2|1)$. For further details on superconformal algebras and their representations, see~\cite{Cordova:2016emh}. These representations are labeled by the the conformal dimension $\Delta$, the $U(1)_R$ charge $r$, and Lorentz spins $(j_1, j_2)$ of the highest weight. Here, $(j_1,j_2)$ are Cartans of $\mathfrak{su}(2)\times\mathfrak{su}(2)\simeq \mathfrak{so}(4)$ and half-integral in our convention (which is different from \cite{Cordova:2016emh}). Including the $SU(2)_F$ flavor spin $k$, also half-integral, we denote the BPS operator (or its representation) as $X\B{Y}[\Delta,j_1,j_2,r]\otimes[k]$, where $X$ and $Y$ specify the type of shortening condition.\footnote{On the supergravity side, the KK spectrum can be organized more systematically by introducing two additional symmetries: $U(1)_P$ and $U(1)_Y$. 
% The first is a subgroup of the Lorentz symmetry $SU(2)_2$(\textbf{JC: then $j_2$...? $SU(2,2|1)$ contains an overall $U(1)$, besides the Lorentz symmetry and R-symmetry. Presumably $U(1)_P$ should be identified with this...?} \textbf{SK: No, it corresponds to charge under $\frac{1}{2}(T_1+T_2)$, while the overall $U(1)$ you mentioned is $T_3$ in our convention discussed at the beginning.} \textbf{JC: I see. Then is it okay to erase the word `Lorentz symmetry $SU(2)_2$?}), while the second is an extra ``bonus'' symmetry, arising from a compact subgroup of the $SL(2,\mathbb{R})$ symmetry of the 5d supergravity. 
% The latter reflects the large-$N$ limit of the $SL(2,\mathbb{Z})$ duality. 
The first is generated by $P\equiv \frac{R_1+R_2}{2}$ (see, for example, \eqref{eq:ruvrir}), and the second is the `bonus' symmetry of supergravity, which is the abelian subgroup of $SL(2,\mathbb{R})$.
A particular linear combination of these charges, $P+2Y$, appears in Table~\ref{tab: Single-trace graviton}, as a superscript on the flavor spin. These symmetries, however, will not play any role in our field-theory analysis, as they are not symmetries of the SCFT.}
For example, $A_1\bar{A}_1[3,\tfrac{1}{2},\tfrac{1}{2},0]$ corresponds to the stress-tensor multiplet, while $A_2\bar{A}_2[2,0,0,0]$ corresponds to the flavor-current multiplet in the IR $\C{N}=1$ SCFT.

\begin{table}
\renewcommand{\arraystretch}{2}
    \centering
    \begin{adjustbox}{max width=\textwidth}
    \begin{tabular}{c|c|c}
        Multiplet & BPS Operator & KK-level \\\hline\hline
        $B_1\B{L}[\frac{9+3n}{4},0,\frac{1}{2},-\frac{n+3}{2}]\otimes[\frac{n+1}{2}]^{(-\frac{n+3}{2})}$ & $\Tr(\lambda_{\alpha} \phi^{(a_1}\cdots \phi^{a_{n+1})})$ & $n\geq0$ \\\hline
        $B_1\B{L}[\frac{6+3n}{4},0,0,-\frac{n+2}{2}]\otimes[\frac{n+2}{2}]^{(-\frac{n+2}{2})}$ & $\Tr(\phi^{(a_1}\cdots \phi^{a_{n+2})})$ & $n\geq0$ \\\hline
        $B_1\B{L}[\frac{12+3n}{4},0,0,-\frac{n+4}{2}]\otimes[\frac{n}{2}]^{(-\frac{n+4}{2})}$ & $\Tr(\lambda^{\alpha}\lambda_{\alpha}\phi^{(a_1}\cdots\phi^{a_n)})$ & $n\geq0$ \\\hline\hline
        $ A_1\B{L}[\frac{6+3n}{2},\frac{1}{2},\frac{1}{2},-n]\otimes[0]^{(0)}$ & $\Tr([(\phi\phi)^n,\lambda_{\alpha}]f-[(\phi\phi)^n,\psi^a]D_{\alpha}\phi_a-[(\phi\phi)^n,\phi^a]D_{\alpha}\psi_{a})$ & $n\geq 1$ \\ 
        $A_1\B{A}_1[3,\frac{1}{2},\frac{1}{2},0]\otimes[0]^{(0)}$ & $\Tr(f\lambda_{\alpha}-\frac{3}{4}\psi_{a} D_{\alpha}\phi^a+\frac{1}{4}\phi^aD_{\alpha}\psi_{a})$ & $n=0$ \\\hline
        $A_1\B{L}[\frac{6+3n}{2},\frac{1}{2},0,-n]\otimes[0]^{(1)}$ & \makecell[c]{$\Tr\big(\lambda^{\alpha}\lambda_{\alpha}f-\psi^a\lambda^{\alpha} D_{\alpha}\phi_a+\frac{1}{2}\lambda^{\alpha}\psi^aD_{\alpha}\phi_a+\frac{1}{2}\phi^a\lambda^{\alpha} D_{\alpha} \psi_{a}$\\$-\phi^a\phi^bD^{\alpha}\phi_a D_{\alpha}\phi_b+2\phi^b(D^{\alpha}\phi^a)\phi_a D_{\alpha}\phi_b\big)$} ($n=1$) & $n\geq 1$ \\\hline
        $A_1\B{L}[\frac{6+3n}{2},\frac{1}{2},0,-n]\otimes[0]^{(-1)}$ & $\Tr(f(\phi\phi)^{n+1}+\frac{1}{4}\sum_{k=0}^n(\phi\phi)^{n-k}\psi^a(\phi\phi)^k\psi_{a})$  & $n\geq0$ \\ 
        % $A_1\B{L}[3,\frac{1}{2},0,0]\otimes[0]^{(-1)}$ & $\Tr((\bar{\phi}\bar{\phi})F_{++}+\frac{1}{4}(\psi_+\psi_+))$ & $n=0$ \\
        \hline
        $ A_2\B{L}[\frac{11+3n}{4},0,\frac{1}{2},-\frac{n+1}{2}]\otimes[\frac{n+1}{2}]^{(-\frac{n+1}{2})}$ & $\Tr(\lambda_{\alpha}\phi^{(a_1}\cdots\phi^{a_n}\psi^{a_{n+1})}-2\phi^{(a_1}\cdots \phi^{a_n}(\phi\phi)D_{\alpha}\phi^{a_{n+1})})$ & $n\geq0$ \\\hline
        $A_2\B{L}[\frac{8+3n}{4},0,0,-\frac{n}{2}]\otimes[\frac{n+2}{2}]^{-(\frac{n}{2})}$ & $\Tr(\phi^{(a_1}\cdots\phi^{a_{n+1}}\psi^{a_{n+2})})$ & $n\geq 1$ \\
        $A_2\B{A}_2[2,0,0,0]\otimes[1]^{(0)}$ & $\Tr(\phi^{(a}\psi^{b)})$ & $n=0$ \\\hline
        $ A_2\B{L}[\frac{14+3n}{4},0,0,-\frac{n+2}{2}]\otimes[\frac{n}{2}]^{(-\frac{n+2}{2})}$ & $\Tr(\lambda^{\alpha}\lambda_{\alpha}\phi^{(a_1}\cdots\phi^{a_{n-1}}\psi^{a_n)}-2\{\lambda^{\alpha},\phi^{(a_1}\cdots\phi^{a_{n-1}}(\phi\phi)\}D_{\alpha}\phi^{a_n)})$ & $n\geq 1$\\ \hline
    \end{tabular}
    \end{adjustbox}
    \caption{Single-trace gravitons from the Kaluza-Klein spectrum. The table lists explicit representatives of BPS graviton operators. For the $B_1\B{L}$-type multiplet, the highest-weight state is itself BPS with respect to the chosen supercharge $Q = Q_-$. For the $A_i\B{L}$-type multiplets ($i = 1,2$), the BPS component is obtained by acting with $Q_+$ on the highest-weight state. From this structure, one can infer the conformal dimension, spins, R-charge, and flavor representation of the BPS operators.}
    \label{tab: Single-trace graviton}
\end{table}

In the $\C{N}=4$ case, the whole single trace graviton multiplet (`$S_n$' \cite{Kinney:2005ej,Choi:2023znd}) was in an \emph{absolutely protected} multiplet $B_1\bar{B_1}[0;0]_{\Delta=n}^{(0,n,0)}$ \cite{Cordova:2016emh}, meaning there is no way to form a long multiplet by combining with other shortend multiplets.
On the other hand, in $\C{N}=1$ superconformal case, only a small fraction of the single trace gravitons are absolutely protected. 
An $\C{N}=1$ superconformal multiplet is absolutely protected when it is either the anti-chiral free field $B_1\bar{A}_l[-\frac{3}{2}r,0,\bar{j},r]$, or anti-chiral $B_1\bar{L}[-\frac{3}{2}r,0,\bar{j},r]$ with $\frac{2}{3}\bar{j}+\frac{2}{3}<-r<\frac{2}{3}\bar{j}+2$ (together with their complex conjugate chiral multiplets).
Therefore, in our case, only following are absolutely protected.
\begin{align}
    \Tr \phi^{(a}\phi^{b)}\;,\;\;\Tr \phi^{(a}\phi^b\phi^{c)}\;,\;\;
    \Tr \lambda_\alpha \phi^a\;,
\end{align}

In the BMN sector, where we restrict ourselves to the fields $(\phi^a, \psi_{a}, f)$, the single-trace gravitons are then given by ($n\geq 0$):
\begin{equation}\label{eqn: Single graviton in BMN}
    \begin{split}
        &\;\;\;u_{n}^{a_1\cdots a_{n+2}}\equiv
        \Tr(\phi^{(a_1}\cdots\phi^{a_{n+2})}),
        \\
        &v_{n}^{a_1\cdots a_{n+2}}\equiv\Tr(\phi^{(a_1}\cdots \phi^{a_{n+1}}\psi^{a_{n+2})}),\\
        w_{n}\equiv&\Tr(f(\phi\phi)^{n+1}+\frac{1}{4}\sum_{k=0}^n(\phi\phi)^{n-k}\psi^a(\phi\phi)^{k}\psi_{a}) \ .
    \end{split}
\end{equation}
Note that $w_n$ is non-Coulomb for $n\geq 1$.
One can regard these as part of the UV BMN gravitons. The precise relation to the UV gravitons will be explained in Seciton \ref{section:IR cohomology in UV coordinates}.

To illustrate how to find the representatives of these graviton cohomologies, let us explain how we obtained $w_n$ which might be the simplest non-trivial example.
From the (BPS) KK spectrum (the left column of Table \ref{tab: Single-trace graviton}), we know that there should exist a tower of operators that are in the representation
$$A_1\B{L}\Big[\frac{6+3n}{2},\frac{1}{2},0,-n\Big]\otimes[0]\;,
\quad (n=0,1,2,\cdots).$$
In fact, there are two towers of the same representation except that one tower starts at $n=1$.
We focus on the tower that starts at $n=0$.
Since the BPS component (with respect to $Q=Q_-$) in the $A_i\B{L}$-type multiplet is obtained by acting $Q_+$ on the highest weight, the BPS operators we want have charges
$$\Delta=\frac{7+3n}{2},\;\;\;j_1=1,\;\;\;j_2=0,\;\;\;r=-n-1,\;\;\;r_F=0,\quad (n\geq 0)$$
% They should consist of the letters $(\phi^a,\psi_a,f)$.
% In fact, there are two towers of operators with the charges above except that one tower starts at $n=1$.
% Therefore, it is reasonable to first focus on $n=0$.
Since the R-charges and $j_1,j_2$ add up, it is easy to identify the letter contents.
From $j=1$, there should be one $f=F_{++}$, or two from either $\psi_a=\psi_{+a}$ or $D_\alpha=D_{+\alpha}$.
From $r=-n-1$ and $j_2=0$, the allowed letter contents are
$$(f\phi^{2n+2}\oplus \psi^2\phi^{2n})\oplus (f\lambda_\alpha\lambda^\alpha \phi^{2n-2}\oplus \lambda^\alpha D_\alpha \psi^1\phi^{2n+1}
\oplus D_\alpha D^\alpha \phi^{2n+2})\;,$$
where we divided into BMN and non-BMN sectors.
Also, since they are flavor-singlets, the flavor indices should all be contracted.

Now we want to find a $Q$-closed \emph{single trace} operator with the letter contents above.
Since there are many ways to contract the flavor indices, one could write down many single trace operators. 
% For $f\phi^{2n+2}$, there are $n+$ flavor singlet single trace operators.
Therefore, one should start with a guess. 
Noting that $n$ starts from $0$ and gradually increases by 1, the most natural candidate for the $f\phi^{2n+2}$ operator would be
$$\Tr f(\phi\phi)^{n+1}\;.$$
The $Q$-action on this is
\begin{align*}
    Q\Tr f(\phi\phi)^{n+1}
    &=-\Tr [\phi^a,\psi_a](\phi\phi)^{n+1}
    =\Tr \psi_a [\phi^a,(\phi\phi)^{n+1}]\\
    &=-\frac{1}{2}\sum_{k=0}^n\Tr\psi_a (\phi\phi)^k Q\psi^a (\phi\phi)^{n-k}\\
    &=-\frac{1}{4}\sum_{k=0}^n\Tr \big(
    (\phi\phi)^kQ\psi^a(\phi\phi)^{n-k}\psi_a+
    (\phi\phi)^{n-k}Q\psi^a(\phi\phi)^{k}\psi_a
    \big)\\
    &=-\frac{1}{4}\sum_{k=0}^n\Tr \big(
    (\phi\phi)^kQ\psi^a(\phi\phi)^{n-k}\psi_a-
    (\phi\phi)^{k}\psi^a(\phi\phi)^{n-k}Q\psi_a
    \big)\\
    &=Q\Big(-\frac{1}{4}\sum_{k=0}^n\Tr (\phi\phi)^k\psi^a(\phi\phi)^{n-k}\psi_a\Big)\;,
\end{align*}
Therefore, we found a $Q$-closed single trace operator, which is precisely $w_n$.

%%%%%%%%%%%%%%%%%%%%%%%%%%%%%%%%%%%%%%%%%
%%%%%%%%%%%%%%%%%%%%%%%%%%%%%%%%%%%%%%%%%
%%%%%%%%%%%%%%%%%%%%%%%%%%%%%%%%%%%%%%%%%
%%%%%%%%%%%%%%%%%%%%%%%%%%%%%%%%%%%%%%%%%
\subsubsection{Non-exactness of $w_n$ and stringy exclusion principle}\label{subsub: nonexact wn}
However, it is not straightforward to show that $w_n$ is not $Q$-exact.
In the $\C{N}=4$ case, all single trace gravitons are of Coulomb type and there was no need to check if they are $Q$-exact.
$w_n$ is non-Coulomb for $n\geq 1$, hence it is nontrivial.
In fact, it is exact when $n\geq N-1$ (analogous to the stringy exclusion principle in $AdS_3$), while it is not exact when $n\leq N-2$.

This is because, for $n\geq N-1$, one can use the Cayley-Hamilton theorem to write $w_{n\geq N-1}$ with multi-traces of $w_{k-2}$ and $\Tr (\phi\phi)^k$ where $2\leq k\leq N-2$.
That is, for any $N\times N$ matrix $M$,
\begin{align}
    M^N=a_{N-1}M^{N-1}+\cdots+a_1 M+a_0 I_N \;,
\end{align}
where the coefficients $a_k$ are some multi-traces of $M$ that are of order $\sim M^{N-k}$. For example, $a_{N-1}=\Tr M$, $a_{N-2}=\frac{1}{2}(\Tr M^2-(\Tr M)^2)$, and $a_0=(-1)^{N+1}\det M$.
More generally, 
\begin{align}\label{eq:Cayley-Hamilton l}
    M^{N+l}=\sum_{k=0}^{N-1} a^{(l)}_kM^k\;,\;\;\;(l\geq 0)
\end{align}
where the coefficients $a_k^{(l)}$ are again some multi-traces of $M$ with powers of $M$ in the trace being no greater than $N$. These are related to $a_k=a_k^{(0)}$ by the following recurrence relation
\begin{align*}
    a^{(l+1)}_k=a_k^{(0)}a_{N-1}^{(l)}+a_{k-1}^{(l)}\;,\;\;\;\; (a_{-1}^{(l)}=0,\;\;\; 0\leq k\leq N-1)\;.
\end{align*}
By acting $P^i\!_j\frac{\partial}{\partial M^i\!_j}$ on the identity \eqref{eq:Cayley-Hamilton l}, we obtain
\begin{equation}
\begin{aligned}
    &\big(PM^{N+l-1}+MPM^{N+l-2}+\cdots+M^{N+l-1}P\big)\\
    &=\sum_{k=1}^{N-1}a_k^{(l)}(PM^{k-1}+\cdots+M^{k-1}P)
    +\sum_{k=0}^{N-1} b_k^{(l)} M^k
\end{aligned}
\end{equation}
where $b^{(l)}_{k}=P^i\!_j\frac{\partial a^{(l)}_k}{\partial M^i\!_j}$. 
% The details of the coefficients $a_k$ and $b_k$ are not important for now. 
% The essence is that any totally symmetrized product of $N$ (same or different) matrices can be expressed in terms of the smaller powers of matrices.
Setting $M=(\phi\phi)$, 
\begin{align}\label{eq:wfNl}
    \Tr f(\phi\phi)^{N+l}=\sum_{k=0}^{N-1} a^{(l)}_{k}\Tr f(\phi\phi)^{k}=\sum_{k=0}^{N-2}a_{k+1}^{(l)}\Tr f (\phi\phi)^{k+1}\;,
\end{align}
Also,
\begin{equation}\label{eq:wpNl}
\begin{aligned}
    &\frac{1}{4}\sum_{k=0}^{N+l-1}\Tr (\phi\phi)^k\psi^a(\phi\phi)^{N+l-1-k}\psi_a\\
    &=-\frac{1}{2}\Tr \big(\psi_1 M^{N+l-1}+M\psi_1 M^{N+l-2}+\cdots +M^{N+l-1}\psi_1\big)\psi_2\\
    &=-\frac{1}{2}\sum_{k=1}^{N-1}a_k^{(l)}\Tr \big(\psi_1 M^{k-1}+\cdots+M^{k-1}\psi_1\big)\psi_2
    -\frac{1}{2}\sum_{k=0}^{N-1}b_k^{(l)}\Tr M^k \psi_2
\end{aligned}
\end{equation}
Combining \eqref{eq:wfNl} and \eqref{eq:wpNl},
\begin{equation}
\begin{aligned}\label{eq:wlargen exact}
    w_{N+l-1}=\sum_{k=0}^{N-2}a_{k+1}^{(l)}w_{k}
    -\frac{1}{2}\sum_{k=0}^{N-1}b_k^{(l)}\Tr M^k \psi_2
\end{aligned}
\end{equation}
From this, we can conclude that $w_{N+l-1}$ is $Q$-exact, where $l\geq 0$.
This is because $a_{k+1}^{(l)}$, $b^{(l)}_k$, and $\Tr M^k\psi_a$ are $Q$-exact, and $w_k$ is $Q$-closed.
$a_{k+1}^{(l)}$ is $Q$-exact because 
$$\Tr (\phi\phi)^{k+1}=-\Tr [X,Y](\phi\phi)^{k}=-\Tr\!\, [Y,(\phi\phi)^{k}]X=-\frac{1}{2}Q\Tr \big(\psi_1(\phi\phi)^{k-1}+\cdots (\phi\phi)^{k-1}\psi_1\big)X\;.$$
and $a_{k+1}^{(l)}$ is a multi-trace of the above operator.
Likewise, 
$$Q\Tr f(\phi\phi)^k\phi_a=\Tr\!\, [\phi_b,\psi^b](\phi\phi)^k\phi_a=\Tr [\phi_a,\phi_b]\psi^b (\phi\phi)^k=\Tr (\phi\phi)\psi_a(\phi\phi)^k=\Tr M^{k+1}\psi_a$$
and $b_k^{(l)}$ is a multi-trace of $\Tr M^k$ with one $\Tr M^l\psi_a$ factor.

%P\frac{\partial c_{N-k}(M)}{\partial M}

%\Big(P^i\!_j\frac{\partial a_k^{(l)}}{\partial M^i\!_j} \Big)

% (\textbf{JC: Prediction: $w_n$ maps to $w_n^{UV}=\Tr (Z^{n+1}f+Z^n\psi_1\psi_2+\cdots)$ in the UV. When $n\geq N-1$, $w_n$ is generated by $\Tr Z^k$ and $w_{k-2}^{UV}$ ($2\leq k\leq N$). But $\Tr Z^k$ is $Q$-exact, making $w_{n\geq N}^{UV}$ $Q$-exact. The exactness of $w^{IR}_{n\geq N-1}$ can presumably be shown similarly.})

Showing the non-exactness of $w_{n\leq N-2}$ is also non-trivial. For $n\leq N-2$, trace relations play no role, and it suffices to show that $w_{n\leq N-2}$ is not \emph{single-trace} exact, i.e. there is no single trace operator that yields $w_n$ by the $Q$-action.
One possible strategy is to demonstrate that $\Tr f(\phi\phi)^{n+1}$ is not single-trace $Q_\psi$-exact, where we denote $Q_\psi$ for the $Q$-action on $\psi_a$ only. If $w_n$ were single-trace exact, $\Tr f(\phi\phi)^{n+1}$ should arise from a term of the form $Q_\psi \Tr f W(\psi,\phi)$, where $W(\psi,\phi)$ is a flavor-singlet word with letter content $\psi\phi^{2n-1}$. This implies that
\begin{align}
    QW(\psi,\phi) = (\phi\phi)^{n+1}
    = (M^{1/2}Z - Q\chi)^{n+1}
    = M^{\frac{n+1}{2}} Z^{n+1} + Q(\cdots).\nonumber
\end{align}
Hence, a necessary condition for $w_n$ to be single-trace exact is that $Z^{n+1}$ is $Q$-exact as a matrix.  
From the flavor $SU(2)$ and R-symmetry constraints, the possible $Q$-preimages must be composed of letters
\begin{align}
    Z^n\psi_3 \oplus Z^{n-1}\phi^a\psi_a.\nonumber
\end{align}
The first letter content yields terms of the form $MZ^{n+1} + Z^{k}[X,Y]Z^{n-k}$. The unwanted contribution $Z^{k}[X,Y]Z^{n-k}$ is equivalent, up to $Q$-exactness, to $Z^{n}[X,Y]$, since
\begin{align}
    -Q[\phi^a,\psi_a] = [Z,[X,Y]].\nonumber
\end{align}
More generally, one may regard $X$ and $Y$ as commuting with $Z$ up to $Q$-exactness, so their positions can be rearranged freely within a trace modulo $Q$-exact terms. However, it is not possible to eliminate the remaining $Z^{n}[X,Y]$ term by any other $Q$-exact matrix, implying that $\Tr f(\phi\phi)^{n+1}$ cannot be single-trace $Q_\psi$-exact.

We also checked the non-exactness of $w_n$ at $N=\infty$ for $n=1,2,3,4,5$ in a brute force method using \texttt{Mathematica}: we constructed a linearly independent basis of $Q$-exact single-trace operators and checked whether the cohomology representative $w_n$ lies within its span. See Table~\ref{tab: Number of Q-exact basis}. Note that $w_0$ is non-exact because it is Coulomb type.

\begin{table}[H]
    \centering
    \begin{tabular}{c|c|c}
      & Before acting $Q$ & $Q$-exact operators \\ \hline
      $n=1$ & $2$ & $1$ \\ \hline
      $n=2$ & $10$& $8$ \\ \hline
      $n=3$ & $40$ & $35$ \\ \hline
      $n=4$ & $147$ & $74$ \\ \hline
      $n=5$ & $526$ &  $300$ \\ \hline
    \end{tabular}
    \caption{Left column: the number of single trace operators. Right column: size of the Q-exact basis which has the same charges as $w_n$.}
    \label{tab: Number of Q-exact basis}
\end{table}

\subsubsection{$SU(2)$ BMN graviton generators}
If we focus on the $SU(2)$ gauge group, the $SU(2)$ adjoint fields $(\phi^a, \psi_a, f)$ can be expressed using the basis of Pauli matrices $\vec{\sigma} = (\sigma_1, \sigma_2, \sigma_3)$, which satisfy the commutation relation $[\sigma_i, \sigma_j] = 2i \epsilon_{ijk} \sigma_k$. We use the following convention.
\begin{equation}\label{eq:so3vec notation}
    \phi^a=\frac{i}{2}\vec{\phi}^a\cdot\Vec{\sigma},
    \qquad\psi_{a}=i\vec{\psi}_a\cdot\Vec{\sigma},\qquad
    f=-i\vec{f}\cdot\Vec{\sigma} \ .
\end{equation}
In vector notation (omitting the $\vec{\;\;\;}$), the $Q$-action in the $SU(2)$ IR theory takes the following expression:
\begin{equation}
    Q\phi^a=0, \qquad Q\psi_a=\phi_a\times(\phi^1\times\phi^2)
    =(\phi_a\cdot\phi^b)\phi_b
    , \qquad Qf=\psi_a\times\phi^a \ ,
\end{equation}
The lowest-level BMN graviton cohomologies are then given as\footnote{Here, we have redefined the operators and absorbed certain normalization factors when taking the trace, so that $u \propto u_0$, $v \propto v_0$, and $w \propto w_0$. However, we will sometimes keep the explicit subscripts $n=0,1,2,\cdots$ when needed, for example, when writing the trace in explicit form.}
\begin{equation}\label{eqn: su2 generating grav}
    \begin{split}
        u^{ab}\equiv\phi^a\cdot\phi^b,\qquad
        v^{ab}\equiv\phi^{(a}\cdot\psi^{b)},\qquad
        w\equiv\phi^1\times\phi^2\cdot f+\psi^1\cdot\psi^2 \ .
    \end{split}
\end{equation}
These gravitons `generate' all BMN graviton cohomologies: that is, all higher-energy graviton cohomologies can be expressed as sums of products of this generating set, modulo $Q$-exact terms.
This can be readily verified for the $u_n$ and $v_n$ gravitons, as they transform in totally symmetric representations, allowing one to focus on their highest-weight components $i.e.$ $u_n^{1\cdots1}$ and $v_n^{1\cdots1}$. For any $2\times 2$ traceless bosonic matrix $M$, $M^2=\frac{1}{2}(\Tr M^2) I_2$. Therefore, $u_{2k+1}=v_{2k+1}=0$ and $u^{1\cdots1}_{2k}=\frac{1}{2^{k}}(u_0^{11})^{k+1}$, $v_{2k}^{1\cdots1}=\frac{1}{2^k}(u_0^{11})^{k}v_0^{11}$.
The nontrivial part is $w_n$. We demonstrate that $w_{2k-1}=0$ and $w_{2k}$ is $Q$-exact for $k\geq1$, by explicitly finding the coefficients in the relation \eqref{eq:wlargen exact} in appendix \ref{minimal check}.

%%%%%%%%%%%%%%%%%%%%%%%%%%%%%%%%%%%%%%%%%
%%%%%%%%%%%%%%%%%%%%%%%%%%%%%%%%%%%%%%%%%
%%%%%%%%%%%%%%%%%%%%%%%%%%%%%%%%%%%%%%%%%
%%%%%%%%%%%%%%%%%%%%%%%%%%%%%%%%%%%%%%%%%
\subsection{Coulomb graviton index}\label{sub:coulomb grav index}
% As discussed earlier, in the IR SCFT the notions of eigenvalues and gravitons no longer coincide, unlike in $\C{N}=4$ SYM. In fact, the eigenvalue method captures only a subset of the graviton spectrum. Indeed, as we will see, there exists a graviton operator that is not detected by the analysis presented in this section.

% (\textbf{JC: we must refer to the previous section on the classification of cohomologies. $\C{N}=4$ gravitons seem to be eigenvalue, whereas only a proper subset of gravitons are eigenvalue in LS. For example, $w_0^2$ for $N\geq 2$, or the whole $w_{n}$ for $N\geq 3$(?)})

As we have explained earlier, gravitons are not necessarily Coulomb type. 
However, it is still useful to define Coulomb type graviton index, following the strategy of \cite{Choi:2023vdm,Choi:2023znd}.
It counts a subset of graviton cohomologies that are of Coulomb type.
By subtracting it from the full index, we obtain the index that receives contributions from non-gravitons and non-Coulomb gravitons.

Let us focus on the BMN sector from now on.
To count linearly independent BMN $SU(2)$ gravitons up to $Q$-exact operators and also up to non-Coulomb gravitons, we rewrite \eqref{eqn: su2 generating grav} using the diagonalized $2\times 2$ traceless matrix fields $(\phi^a, \psi_a, f)$. For example, take $X = \text{diag}(x, -x)$, and similarly for the others. The generating set is then given by
\begin{equation}\label{BMN su2 graviton eigenvalue}
    \begin{split}
        \begin{split}
            &x^2,\qquad y^2,\qquad xy,\\
            &x\psi^1,\qquad y\psi^2,\qquad x\psi^2+ y\psi^1,\\
            &\psi^1\psi^2 \ .
        \end{split}
    \end{split}
\end{equation}
Counting these Coulomb gravitons can be carried out analytically, as performed in \cite{Choi:2023znd}. With a slight abuse of notation, we assign fugacity $x,y$ to each eigenvalue $x,y$. For fermions, we assign $x^2y\eta$ to $\psi^1$ and $xy^2\eta$ to $\psi^2$ where $\eta$ is a complex number. Since there are only two fermionic components, we can divide (multi-)gravitons into 0-fermion, 1-fermion, and 2-fermion sectors. The set of 0-fermion gravitons is
\begin{align}
    S_{0f}=\{x^{a}y^{b}|a,b\geq 0, (a+b)\text{ even}\}.
\end{align}
The generating function for $S_{0f}$ can be obtained using the following trick:
\begin{equation}
    Z_{0f}=\frac{1}{2}\bigg(\sum_{a,b}x^ay^b+\sum_{a,b}(-1)^{a+b}x^ay^b\bigg)=\frac{1+xy}{(1-x^2)(1-y^2)}\ .
\end{equation}
We sum over all powers of $x$ and $y$, then project onto the Weyl-invariant sector, $i.e.$ under the identifications $x \to -x$ and $y \to -y$.

The 1-fermion sector is slightly more complicated but remains tractable. We first include multi-gravitons obtained by multiplying scalar gravitons $(x^2,y^2,xy)$ with $x\psi^1$ and $y\psi^2$. We then consider those formed with $x\psi^2 + y\psi^1$. Among these, terms proportional to $xy(x\psi^2 + y\psi^1)$ are already included, as they can be expressed as linear combinations of $xy \cdot x\psi^1$ and $xy \cdot y\psi^2$. As a result, we only need to account for the gravitons of the form $x^{2k}(x\psi^2+y\psi^1)$ and $y^{2k}(x\psi^2+y\psi^1)$ where $k=0,1,2,\cdots$. In total, the set of 1-fermion gravitons is
\begin{align}
    S_{1f}=&\{g x\psi^1\;|\;g\in S_{0f}\}\cup 
    \{g y\psi^2\;|\;g\in S_{0f}\}\nonumber\\
    &\cup
    \{x^{2k}(x\psi^2+y\psi^1)\;|\;k=0,1,2,\cdots\}
    \cup\{y^{2k}(x\psi^2+y\psi^1)\;|\;k=1,2,3,\cdots\} \ .
\end{align}
Here, note that all the sets are disjoint. For example, the last set does not contain $(x\psi^2+y\psi^1)$ itself.
Therefore, the generating function for $S_{1f}$ is
\begin{equation}
    \begin{split}
        Z_{1f}&=Z_{0f}x^3y\eta+Z_{0f}xy^2\eta+\frac{x^2y^2\eta}{1-x^2}+\frac{x^2y^2\eta}{1-y^2}-x^2y^2\eta\\
        &=\frac{xy\eta(x^2+y^2+xy+x^3y+xy^3-x^3y^3)}{(1-x^2)(1-y^2)} \ .
    \end{split}
\end{equation}
The 2-fermion sector is simply
\begin{align}
    Z_{2f}=Z_{0f}x^3y^3\eta^2 \ .
\end{align}
Therefore, BMN $SU(2)$ Coulomb graviton generating function is
\begin{align}
    Z^{SU(2)}_{\TR{Coulomb grav}}
    =\frac{1+xy}{(1-x^2)(1-y^2)}
    \big[
    1+xy\eta(x^2+y^2+xy-x^2y^2)+x^3y^3\eta^2
    \big] \ .
\end{align}
The BMN Coulomb graviton index is given by setting $\eta=-1$
\begin{align}
    \C{I}_{\TR{Coulomb grav}}^{SU(2)}
    =\frac{1+xy}{(1-x^2)(1-y^2)}
    \big[
    1-xy(x^2+y^2+xy-2x^2y^2)
    \big] \ .
\end{align}
where $x=t^{3/2}s$, $y=t^{3/2}s^{-1}$. Let us define $q\equiv t^{3/2}$. Note that when we set $s = 1$, the BMN Coulomb graviton index becomes a polynomial.
\begin{align}
    \C{I}_{\TR{Coulomb grav}}^{SU(2)}=1+3q^2+2q^4.
\end{align}

On the other hand, the full BMN index is
\begin{equation}
    \begin{split}
        \C{I}^{SU(2)}&=
        \frac{(1-x^2y)(1-xy^2)P(x,y)}{(1-x^2)(1-y^2)(1-x^4y^4)(1-x^3y^2)(1-x^2y^3)},
    \end{split}
\end{equation}
where
\begin{equation}
    \begin{split}
        P(x,y)&=1+xy+xy(x+y)-xy(x^2+y^2+xy)+2x^3y^3-2x^3y^3(x+y)\\
        &\quad+x^4y^4(x+y)-x^4y^4(x^2+y^2+3xy)+x^5y^5(x^2+y^2+2xy)\\
        &\quad-x^6x^6(x+y)+x^7y^7(x+y)\ .
    \end{split}
\end{equation}
Its unrefined version, expressed in terms of $q \equiv t^{3/2}$ with $s = 1$, is
\begin{align}
    \C{I}^{SU(2)}&=
    \frac{(1-q^3)^2}{(1-q^5)^2(1-q^8)}
    \Big[
    1+3q^2+2q^3+2q^4+4q^5+3q^6+2q^7+4q^8+2q^9+2q^{11}
    \Big]\nonumber\\
    &=1+3q^2+2q^4+4q^7+2q^9+\cdots \ .
\end{align}

%%%%%%%%%%%%%%%%%%%%%%%%%%%%%%%%%%%%%%%%%
%%%%%%%%%%%%%%%%%%%%%%%%%%%%%%%%%%%%%%%%%
%%%%%%%%%%%%%%%%%%%%%%%%%%%%%%%%%%%%%%%%%
%%%%%%%%%%%%%%%%%%%%%%%%%%%%%%%%%%%%%%%%%
\subsection{Coulomb non-graviton cohomologies}\label{sub: Coulomb nongrav}
We define the non-(Coulomb graviton) index:
\begin{equation}
    \C{I}_{\TR{non-CG}}^{SU(2)}=\C{I}^{SU(2)}-\C{I}_{\TR{Coulomb grav}}^{SU(2)}\ ,
\end{equation}
which contains information about non-gravitons and non-Coulomb gravitons. 
Using  $x=q s$ and $y=q s^{-1}$, it can be expanded as follows:
\begin{align}
    \C{I}_{\text{non-CG}}^{SU(2)}&=
    q^7\chi_3
    +q^9(\chi_5-\chi_3)
    +q^{11}(\chi_7-\chi_5-\chi_3)
    +q^{12}(\chi_4+1)
    \nonumber\\
    &+q^{13}(\chi_9-\chi_7-\chi_5+\chi_3)
    +q^{14}(\chi_6-\chi_4)
    +q^{15}(\chi_{11}-\chi_{9}-\chi_{7}+\chi_5)\nonumber\\
    &+q^{16}(\chi_8-\chi_6-\chi_4-1)
    +q^{17}(\chi_{13}-\chi_{11}-\chi_{9}+\chi_{7}+\chi_5+\chi_1)\\
    &+q^{18}(\chi_{10}-\chi_8-\chi_6+\chi_4+1)+q^{19}(\chi_{15}-\chi_{13}-\chi_{11}+\chi_9+\chi_7-\chi_5-\chi_1)\nonumber
    \\
    &+q^{20}(\chi_{12}-\chi_{10}-\chi_{8}+\chi_6+1)
    +q^{21}(\chi_{17}-\chi_{15}-\chi_{13}+\chi_{11}+\chi_9-\chi_{7}-\chi_5-\chi_1)
    \nonumber
    \\
    &+O(q^{20})\nonumber\ .
\end{align}
where we have defined the flavor $SU(2)$ character $\chi_n$ of the spin $n/2$ representation as
\begin{align}
    \chi_n\equiv \frac{1}{q^n}\sum_{a=0}^n x^{n-a}y^a
    =\sum_{a=0}^n s^{n-2a}
    \ .
\end{align}
Here, $n$ can be thought of as the number of flavor (upper) indices.

%%%%%%%%%%%%%%%%%%%%%%%%%%%%%%%%%%%%%%%%%
%%%%%%%%%%%%%%%%%%%%%%%%%%%%%%%%%%%%%%%%%
\subsubsection{$q^7\chi_3$ (+`hairy' operators)}
Let us consider the first term in the non-CG(Coulomb graviton) index, $q^{7}\chi_3$. There are only two gauge-invariant operators.
\begin{equation}
    \phi^{(a}\cdot\phi^{b}\phi^{c)}\cdot f,\qquad \phi^{(a}\cdot\psi^b\times\psi^{c)}\ .
\end{equation}
These give rise to the following cohomology:
\begin{align}
    O_7^{abc}=\phi^{(a}\cdot\phi^b\phi^{c)}\cdot f
    -\frac{1}{2}\phi^{(a}\cdot\psi^b\times\psi^{c)} \ .
\end{align}
One can verify that this operator is $Q$-closed as follows. Using $Q\psi^a=(\phi^a\cdot\phi^b)\phi_b$,
\begin{align}
    Q(\psi^b\times\psi^c)
    =2(Q\psi^{(b})\times\psi^{c)}
    =2\phi_d\times\psi^{(b}
    \phi^{c)}\cdot\phi^d \ .
\end{align}
Therefore,
\begin{equation}
    Q\big[\phi^{(a}\cdot(\psi^b\times\psi^{c)})\big]=2\phi^{(a}\cdot\phi_d\times\psi^{b}\phi^{c)}\cdot\phi^d=-2(\phi^1\times\phi^2)\cdot\psi^{(a}\phi^b\cdot\phi^{c)}
\end{equation}
where we have used $\phi^b\times\phi_d=-\delta^b_d(\phi^1\times\phi^2)$. On the other hand, one can show that
\begin{equation}
    Q(\phi^{(a}\cdot\phi^b\phi^{c)}\cdot f)=\phi^{(a}\cdot\phi^b\phi^{c)}\cdot(\psi_d\times\phi^d)=+\frac{1}{2}Q\big[\phi^{(a}\cdot(\psi^b\times\psi^{c)})\big]
\end{equation}
Combining these two equations yields
\begin{equation}
    Q\Big[\phi^{(a}\cdot\phi^b\phi^{c)}\cdot f-\frac{1}{2}\phi^{(a}\cdot (\psi^b\times\psi^{c)})\Big]=0
\end{equation}

The operator $O_7^{(abc)}$ cannot be $Q$-exact, as it is not possible to construct a flavor-spin-$\frac{3}{2}$ operator from the letter content $\psi^a f$, and the $Q$-action preserves the flavor representation. Or simply because it is Coulomb.
Moreover, this operator cannot be a graviton, since no graviton appears at order $q^7$. More generally, no operator at odd powers of $q$ can be a graviton, since all BMN gravitons have even power in $q$: $u\to q^2\chi_2$, $v\to -q^4\chi_2$, $w\to q^6$.

In general $N$, the operator $O_7^{abc}$ can be written either as
\begin{equation}\label{eqn: double trace case}
    \Tr(f\phi^{(a})\Tr(\phi^b\phi^{c)})+\frac{1}{4}\Tr(\psi^{(a}\psi^b\phi^{c)})\ ,
\end{equation}
or as
\begin{equation}\label{eqn: single trace case}
    \Tr(f\phi^{(a}\phi^b\phi^{c)})+\frac{1}{8}\Tr(\psi^{(a}\psi^b\phi^{c)})\ .
\end{equation}
However, in both cases the operators fail to be $Q$-closed for any $N \geq 3$. Namely, the cohomology $O_7^{abc}$ is fortuitous \cite{Chang:2024zqi}.

In fact, the operator $O_7^{abc}$ generates an infinite tower of non-graviton operators contributing to the terms $q^{7+2k} \chi_{2k+3}$ for $k = 1, 2, 3, \ldots$ in the non-CG index. These are constructed by dressing $O_7^{abc}$ with the graviton $u^{ab}$ as
\begin{align}
    O^{a_1\cdots a_{2k+3}}_{7,u^{k}}
    \equiv u^{(a_1a_2}\cdots u^{a_{2k-1}a_{2k}}
    O_7^{a_{2k+1} a_{2k+2} a_{2k+3})}\ ,
\end{align}
This operator transforms in the flavor-spin $\frac{2k+3}{2}$ representation, and remains $Q$-closed due to the Leibniz rule of the $Q$ action. This operator cannot be $Q$-exact, as there is no way to build a flavor-spin $\frac{2k+3}{2}$ operator from $(\phi^a)^{2k} \psi^b f$ --- the number of flavor indices is not enough. 

Furthermore, the series of terms $-q^{11+2k} \chi_{5+2k}$ for $k = 0, 1, 2, \ldots$ is accounted for by the following tower of fermionic operators:
\begin{align}\label{eqn: O_7uv}
    O^{a_1\cdots a_{2k+5}}_{7,u^k v}\equiv
    v^{(a_1a_2}O^{a_3\cdots a_{2k+5})}_{7,u^{k}}.
\end{align}
This operator transforms in the flavor-spin $\frac{2k+5}{2}$ representation and is $Q$-closed. It is built of the letters $(\phi^a)^{2k+4}\psi^b f\oplus (\phi^a)^{2k+2}(\psi^b)^{3}$, and cannot be $Q$-exact, since there is no combination of letters like $(\phi^a)^{2k+1}(\psi^b)^2f \oplus (\phi^a)^{2k+3}f^2$ that yields the required flavor-spin $\frac{2k+5}{2}$. Moreover, the operator has an odd power of $q$, and thus cannot correspond to a graviton.

At order $q^{13+2k}\chi_{2k+3}$, the operator
\begin{align}
    O_{7,u^kw}^{a_1\cdots a_{2k+3}}
    \equiv w O^{a_1\cdots a_{2k+3}}_{7,u^k},
\end{align}
is also not $Q$-exact, as it is of Coulomb type, $i.e.$ non-vanishing when only the third components of the fields are left in the $SO(3)$ vector notation. Since all $Q$-exact operators are of non-Coulomb type, this confirms the non-exactness $O_{7,u^kw}$.

The `dressed' operators, $O_{7,u^k}, O_{7,u^kv}, O_{7,u^kw}$ might be viewed as `hairy' operators \cite{Choi:2023znd}, borrowing terminology from the gravity literature about hairy black holes.
Also, all these cohomology classes constructed here are of Coulomb type.

There could be other hairy operators, and we will address this systematically in Section \ref{sub:nonCoulombnongraviton}.

%%%%%%%%%%%%%%%%%%%%%%%%%%%%%%%%%%%%%%%%%
%%%%%%%%%%%%%%%%%%%%%%%%%%%%%%%%%%%%%%%%%
\subsubsection{$-q^{9}\chi_3$ (+`hairy' operators)}
Let us next consider the term $-q^9\chi_3$. There are only 4 gauge-invariant operators in this sector:
\begin{equation}
    \psi^{(a}\cdot\psi^b\times\psi^{c)},\quad 
    \phi^{(a}\cdot \phi^{b}\psi^{c)}\cdot f,\quad
    \phi^{(a}\cdot \psi^{b}\phi^{c)}\cdot f
    =v^{(ab}\phi^{c)}\cdot f,\quad
    \phi^{(a}\cdot \phi^{b} \phi^{c)}\cdot \phi^{d}
    \psi_{d}\cdot\phi^{1}\times\phi^{2}.
\end{equation}
But the last one is $Q$-exact because $Q(\phi^a\cdot f)=-\psi^a\cdot\phi^1\times\phi^2$.
Using the first three operators, there is only one $Q$-closed operator:
\begin{align}
    O_9^{abc}\equiv\phi^{(a}\cdot\psi^{b}\phi^{c)}\cdot f
    -\frac{1}{6}\psi^{(a}\cdot\psi^{b}\times\psi^{c)}.
\end{align}
When showing the $Q$-closedness, it might be helpful to use the identity $\psi^{(b}_i\psi^{c)}_j=\frac{1}{2}\epsilon_{ijk}(\psi^{(b}\times\psi^{c)})_k$. This operator is manifestly non-exact, as the $Q$-action strictly increases the number of scalar fields. 
Also, this operator cannot be a graviton, as there is no graviton at $q^{9}$.

One can also construct a tower of non-graviton cohomologies contributing to the index as $-q^{11+2k}\chi_{5+2k}$ by considering
\begin{align}
    O_{9,u^{k+1}}^{a_1\cdots a_{2k+5}}\equiv u^{(a_1a_2}\cdots u^{a_{2k+1}a_{2k+2}}O_9^{a_{2k+3}a_{2k+4}a_{2k+5})}.
\end{align}
This tower contributes to the non-graviton part of the index in the same manner as \eqref{eqn: O_7uv} discussed earlier. However, it can be shown that
\begin{align}
    O_{9,u^{k+1}}^{a_1\cdots a_{2k+5}}
    -
    O_{7,u^kv}^{a_1\cdots a_{2k+5}}=0.
\end{align}
$i.e.$ they are identical. For example, at $k=0$, we have
\begin{equation}
    u^{(a_1a_2}O_{9}^{a_3a_4a_5)}
    -
    v^{(a_1a_2}O_{7}^{a_3a_4a_5)}=
    \frac{1}{2}\phi^{(a_1}\cdot \psi^{a_2}
    \phi^{a_3}\cdot\psi^{a_4}\times\psi^{a_5)}
    -\frac{1}{6}\phi^{(a_1}\cdot \phi^{a_2}
    \psi^{a_3}\cdot\psi^{a_4}\times\psi^{a_5)} \ ,
\end{equation}
which vanishes due to the identity $\psi^{(a}_i\psi^b_j\psi^{c)}_k=\frac{1}{6}\epsilon_{ijk}\psi^{(a}\cdot\psi^b\times\psi^{c)}$. Then, cases with $k\geq 1$ follow straightforwardly.

Another way to dress $O_9^{abc}$ with $u^{ab}$ is
\begin{align}
    O^{abc}_{11}\equiv \tensor{u}{^{(a}_d} O^{bc)d}_9,
\end{align}
where $\tensor{u}{^a_b}=\phi^a\cdot\phi_b$. This operator is not $Q$-exact, as it is of Coulomb type. One can further dress it with powers of $u^{ab}$ as
\begin{align}
    O^{a_1\cdots a_{2k+3}}_{11,u^k}
    \equiv u^{(a_1a_2}\cdots u^{a_{2k-1}a_{2k}}O_{11}^{a_{2k+1}a_{2k+2}a_{2k+3})}.
\end{align}
which contributes at $-q^{11+2k} \chi_{3+2k}$ and is again non-exact.

%%%%%%%%%%%%%%%%%%%%%%%%%%%%%%%%%%%%%%%%%
%%%%%%%%%%%%%%%%%%%%%%%%%%%%%%%%%%%%%%%%%
\subsubsection{$q^{12}\chi_4$ (+`hairy' operators)}
In this sector, one can write down 5 gauge-invariant operators.
\begin{equation}
\begin{split}
    \phi^{(a}\cdot\phi^b \phi^c\cdot\phi^{d)}f\cdot f\;,
    \;\;
    &\phi^{(a}\cdot\phi^b \phi^c\cdot f\phi^{d)}\cdot f\;,
    \;\;\\
    \phi^{(a}\cdot\phi^b f\cdot\psi^c\times\psi^{d)}\;,
    \;\;
    \phi^{(a}\cdot\psi^b f\cdot&\phi^c\times\psi^{d)}\;,
    \;\;
    \phi^{(a}\cdot f \phi^b\cdot \psi^c\times\psi^{d)}\;,
\end{split}
\end{equation}
It turns out that the last three operators are linearly dependent. 
\begin{align}\label{eq:relq12chi4}
    \phi^{(a}\cdot\phi^b f\cdot\psi^c\times\psi^{d)}
    -2
    \phi^{(a}\cdot\psi^b f\cdot&\phi^c\times\psi^{d)}
    -
    \phi^{(a}\cdot f \phi^b\cdot \psi^c\times\psi^{d)}=0
    \;,
\end{align}
Therefore, there are 4 independent gauge-invariant operators. Since there is no operator at $\chi_4$ with letters $\phi^a\psi^bf^2$, all these 4 operators are not $Q$-exact.

Up to the relation \eqref{eq:relq12chi4}, there is only one $Q$-closed combination:
\begin{align}
    O_{12}^{abcd}\equiv
    \phi^{(a}\cdot\phi^b \phi^c\cdot f\phi^{d)}\cdot f
    -
    \phi^{(a}\cdot f \phi^b\cdot \psi^c\times\psi^{d)}
    \;.
\end{align}
This operator is again of Coulomb type.
% This operator remains non-vanishing when the fields are diagonal (or equivalently, aligned along the 3-direction in the $SO(3)$ vector language). 
This implies that by dressing it with other Coulomb gravitons, or even with $O_7$ or $O_9$, which are Coulomb non-gravitons, one can easily construct new, non-trivial cohomologies. 
Constructing them explicitly, as we did for $O_7,O_9$, would be interesting, but we can do this more systematically using the Gr\"obner basis method \cite{Choi:2023vdm}.

%%%%%%%%%%%%%%%%%%%%%%%%%%%%%%%%%%%%%%%%%
%%%%%%%%%%%%%%%%%%%%%%%%%%%%%%%%%%%%%%%%%
%%%%%%%%%%%%%%%%%%%%%%%%%%%%%%%%%%%%%%%%%
%%%%%%%%%%%%%%%%%%%%%%%%%%%%%%%%%%%%%%%%%
\subsection{Counting Coulomb cohomologies including non-gravitons}\label{sub:nonCoulombnongraviton}
As mentioned earlier, Coulomb cohomologies can be multiplied to each other to construct new Coulomb cohomologies, as they remain $Q$-closed and non-exact unless there are too many fermions to make the product vanish on the diagonal field configurations. Of course, even if an operator vanishes for diagonal fields, it may still be non-trivial.

Since we have found Coulomb-type non-graviton, it is easy to construct `hairy operators' using such non-gravitons, as we have done so far.
It is even easy to construct (non-graviton)$\times$(non-graviton) cohomology in this way.
In this subsection, we efficiently count these Coulomb type cohomologies that we have found so far and subtract their contributions from the full BMN index.

% \textbf{(JC:expression for the partition function instead of the index? - in my old laptop - I should check it later...)}

Including $O_7,O_9,O_{12}$ beside the gravitons $u,v,w$, the Coulomb-index can be obtained analytically using Gr\"obner basis of the ideal of relations of Coulomb-type cohomologies \cite{Choi:2023vdm}.
In fact, the full set of independent (up to $Q$-exact and non-Coulomb cohomologies) Coulomb cohomologies (that are polynomials of $u,v,w,O_7,O_9,O_{12}$) can be identified using the Gr\"obner basis.
Here, we present the index of Coulomb type cohomologies generated by $(u,v,w,O_7,O_9,O_{12})$.
\begin{equation}
\begin{split}
    \C{I}_{\text{Coulomb},O_{12}}
    =\frac{(1-x^2y)(1-xy^2)P_C(x,y)}{(1-x^2)(1-y^2)(1-x^8y^4)(1-x^4y^8)},
\end{split}
\end{equation}
where (using $x=q s,\;y=qs^{-1}$)
\begin{equation}
\begin{split}
    P_C(x,y)&=1+q^2+q^3\chi_1-q^4\chi_2+q^5\chi_1+2q^6-q^7\chi_1
    +q^8(\chi_2+2)-q^9\chi_3-q^{10}\\
    &+3q^{11}\chi_1-q^{12}(\chi_4+2)+q^{13}(\chi_3+\chi_1)
    +q^{14}(-\chi_4+2\chi_2+1)-q^{15}(\chi_5+3\chi_1)\\
    &+q^{16}(2\chi_2+1)+q^{17}(-\chi_5+\chi_3)-2q^{18}
    +q^{19}(\chi_3+2\chi_1)-q^{20}(\chi_6+\chi_4+1)\\
    &+q^{22}(\chi_4+\chi_2+1)-q^{23}(\chi_5+\chi_3)+2q^{24}
    +q^{25}(\chi_3+\chi_1)+q^{26}(-\chi_4-\chi_2+1)\\
    &+q^{27}\chi_1+q^{28}(\chi_2+1)-q^{29}\chi_3+q^{31}\chi_1
    \;\;.
\end{split}
\end{equation}
Then, the following is the index of BMN cohomologies that are not Coulomb cohomologies generated by $(u,v,w,O_7,O_9,O_{12})$.
\begin{equation}
    \begin{split}
        \C{I}^{SU(2)}_{\mathrm{BMN}}
        -\C{I}_{\text{Coulomb},O_{12}}
        &=q^{12}-q^{14}\chi_4
        -q^{16}+q^{17}(\chi_5+\chi_1)+q^{18}\\
        &\quad-q^{19}(\chi_5+\chi_1)+q^{20}-q^{21}\chi_1
        +q^{22}(\chi_6+\chi_2)\\
        &\quad+q^{23}(\chi_5+\chi_1)
        -q^{24}(\chi_6+\chi_2+1)
        +q^{25}\\
        &\quad+q^{26}(-\chi_8-\chi_6-\chi_2+1)
        +q^{27}(\chi_7+\chi_3-\chi_1)\\
        &\quad+q^{28}(\chi_6+\chi_2+1)
        +q^{29}(\chi_9-\chi_7-\chi_3-\chi_1)
        +q^{30}\chi_2\\
        &\quad+O(q^{31})\;\;.
    \end{split}
\end{equation}
The first term, $q^{12}$, corresponds to the cohomology
\begin{equation}
    w^2=(\phi^1\times\phi^2\cdot f+\psi^1\cdot\psi^2)^2\ ,
\end{equation}
which is not $Q$-exact due to the presence of the term $(\psi^1\cdot\psi^2)^2$, since the $Q$-action strictly increases the number of scalar fields. 
Note that, although the graviton operator $w$ itself does not vanish for diagonal field configurations, its square does. 
Similarly, $q^{18}$ arises from the operator $w^3$, which is not $Q$-exact due to the term $(\psi^1\cdot \psi^2)^3$.

Proceeding further, $-q^{14}\chi_4$ is from
\begin{equation}
    O_{14}^{abcd}=
    f\cdot\phi^{(a}f\cdot\phi^{b}
    \psi^{c}\cdot\phi^{d)}-f\cdot\psi^{(a}\psi^b\cdot\psi^c\times\phi^{d)}\ ,
\end{equation}
which remains non-vanishing for diagonal field configurations and is therefore not $Q$-exact.
Non-exactness can also be easily confirmed from the fact that there is no operator at $\chi_4$ with letter content $(\psi^a)^2f^2\oplus (\phi^a)^2f^3$.
Also, this can't be a graviton, because there is no way to produce $-q^{14}\chi_4$ from $w^2u^{ab}$, which is the only possible combination that produces the $f^2$ term.
Since both the terms of $O_{14}$ can never arise from $Q$-exact operators, $O_{14}$ is non-graviton.
% (\textbf{JC: checked that this is Q-closed.})

Finally, the term $-q^{16}$ is generated by the following cohomology:
\begin{equation}\label{eq: O16}
    \begin{split}
        O_{16}&=\frac{1}{3}(\psi^a\cdot\psi^b\times\psi^c)(\psi_a\cdot\psi_b\times\phi_c)+(f\cdot\phi^a\times\phi_a)(f\cdot\phi^b)(\psi_b\cdot\phi^c\times\phi_c)\\
        &\qquad+(f\cdot\phi^a)(\psi_a\cdot\phi^b\times\phi_b)(\psi^c\cdot\psi_c)+2(f\cdot\psi^a)(\phi_a\cdot\psi^b)(\psi_b\cdot\phi^c\times\phi_c)
    \end{split}
\end{equation}
% (\textbf{JC: 2nd and 3rd term combined is proportional to $w$. maybe not so important...})
% (\textbf{JC: 1st term can be written as $-4(\psi^1\cdot\psi^2)^2(\psi^a\cdot\phi_a)$.})
% (\textbf{JC: checked Q-closedness.})
The first term $\phi \psi^5$ obstructs this cohomology from being $Q$-exact; there is no singlet made of $\psi^4 f$. The gauge invariants are either $(\psi^a\cdot f) (\psi^b\cdot\psi^c\times\psi^d)$ or $(\psi^a\cdot\psi^b)(f\cdot\psi^c\times\psi^d)$ but there is no way to form a flavor singlet. Moreover, it is not possible to construct a singlet with field content $\phi \psi^5$ from gravitons, because it can only appear in $w^2 v^{ab}$. Since the $\phi\psi^5$ singlet term can never originate from a $Q$-exact operator, this proves that $O_{16}$ is not a graviton.
This operator vanishes for diagonal field configurations, making it the first example of non-eigenvalue, non-graviton cohomology, analogous to the fortuitous cohomologies found in the $\mathcal{N}=4$ SYM case.

%%%%%%%%%%%%%%%%%%%%%%%%%%%%%%%%%%%%%%%%%
%%%%%%%%%%%%%%%%%%%%%%%%%%%%%%%%%%%%%%%%%
%%%%%%%%%%%%%%%%%%%%%%%%%%%%%%%%%%%%%%%%%
%%%%%%%%%%%%%%%%%%%%%%%%%%%%%%%%%%%%%%%%%
%%%%%%%%%%%%%%%%%%%%%%%%%%%%%%%%%%%%%%%%%
%%%%%%%%%%%%%%%%%%%%%%%%%%%%%%%%%%%%%%%%%
\section{UV origins of IR cohomologies}\label{section:IR cohomology in UV coordinates}
% In this section, we attempt to match the cohomologies obtained from the analysis of the IR SCFT with those of the $\mathcal{N}=4$ SYM. Strictly speaking, the matching is not directly between the IR and $\mathcal{N}=4$ cohomologies, but rather with the mass-deformed UV cohomologies. Nevertheless, in an appropriate limit where $M \to 0$ can be taken, one may expect to recover a relation to the $\mathcal{N}=4$ SYM case, suggesting that the cohomologies found in the IR could originate from the UV cohomologies.
In this section, we identify the UV origins of `IR' cohomologies that we have found so far in the `IR coordinates'.\footnote{
It is non-trivial to show that every mass-deformed cohomology originates from a UV ($\mathcal{N}=4$) cohomology. This can be inferred from the `homotopy perturbation lemma' of \cite{Budzik:2023xbr}. In Appendix D of \cite{Budzik:2023xbr}, it is shown that for a supercharge $Q=Q_0+\hbar Q_1$, where $\hbar$ is small, there exists an isomorphism between $H(X,Q_0)=:U_0$ and $H(U_0,\tilde{Q}_1)$.
Here, $X$ is the vector space of gauge invariant operators, and $H(X,Q_0)$ is the cohomology with respect to the differential $Q_0$ acting on $X$. $\tilde{Q}_1$ can be expressed as a perturbative deformation of $Q_1$. The important thing is that the cohomology of a differential $\tilde{Q}_1$ acting on $U_0$ is a linear subspace of $U_0$.
}
We express the IR cohomologies in the UV coordinates to find the corresponding UV ($M=0$) cohomologies.
Let us reproduce the field redefinitions \eqref{eqn: UV to IR transf} here, omitting the $UV$ subscript on the right-hand side for simplicity.
\begin{equation}
    \begin{split}\label{eqn: UV to IR transf2}
        {\phi}^a_{IR}&\equiv 
        M^{-1/4}{\phi}^a\ ,\\
        \psi_{IR\,a}&\equiv 2M^{1/4}\Big(\psi_{a}+\frac{1}{M}[{\phi}_{a},\psi_{3}]\Big)\ ,\\
        f_{IR}&\equiv -2\Big(f-\frac{1}{M}\psi_{3}\psi_{3}\Big)\ .
    \end{split}
\end{equation}

In the UV language, it is convenient to decompose the $Q$-action as follows:
\begin{equation}\label{eq: Q=Q0+Q1}
    Q=Q_0+MQ_1\;,
\end{equation}
where $Q_0 = Q_-^4$ is the chosen supercharge in $\C{N}=4$ SYM ($M=0$ version of eq.\eqref{eq: UV Q}), and $Q_1$ is defined as
\begin{equation}
    Q_1=Z\frac{\partial}{\partial\psi_3}
    =Z^i\!_j\frac{\partial}{\partial (\psi_3)^i\!_j}\;.
\end{equation}
When there are derivatives, $Q_1$ also acts on $(D_\alpha)^n\psi_3$ to yield $(D_\alpha)^n Z$, but we will suppress them because we are working in the BMN sector.
% Here, $Q_1$ has mass dimension $-1/2$, so the combination $M Q_1$ has mass dimension $1/2$.

If we plug \eqref{eqn: UV to IR transf2} into an IR cohomology $O_{IR}$, it takes the following form.
\begin{align}\label{eq: IR op in UV}
    O_{IR}=M^{-k-\frac{l}{4}}(O^{(0)}+MO^{(1)}+
    \cdots+M^{m}O^{(m)})\;,
\end{align}
where $k,l$ are typically non-negative integers, and $O^{(i)}$ is a UV operator.
And this expression is ambiguous up to $Q$-exactness; $Q\Lambda=Q_0\Lambda+MQ_1\Lambda$ for some $\Lambda$.
Therefore, naively taking the small $M$ limit and taking the leading operator $O^{(0)}$ does not necessarily give rise to the UV cohomology that originated.

As explained around Table \ref{tab: spurious charges}, if we assign a UV R-charge $-2/3$ to $M$, IR and UV R-symmetries act identically for the IR letters. Therefore, the UV R-charge of the right hand side of \eqref{eq: IR op in UV} is the same as the IR R-charge of $O_{IR}$. Then, one can infer the UV R-charges of UV operators $O^{(i)}$.
$$r^{UV}[O^{(i)}]=r^{IR}[O_{IR}]-r^{UV}[M^{i-k-\frac{l}{4}}]
=r^{IR}[O_{IR}]+\frac{2}{3}\Big(i-k-\frac{l}{4}\Big)\;.$$
Since $O^{(i)}$'s have different R-charges, it is not straightforward to find the UV R-charge of the UV cohomology that is non-exact.
Furthermore, since the $Q$-action produces a power of $M$, the ambiguity of $Q$-exactness is not order-by-order in $M$.
This makes the identification of UV cohomology, to which the IR cohomology reduces in the small mass limit, highly non-trivial in general.

\subsection{Gravitons}
% (\textbf{JC: Other gravitons e.g. $u_n,v_n, w_{n\geq 1}$})
Let us first consider the UV limit of graviton cohomologies.
The UV limit of $u_{IR,n}$ is easy.
\begin{equation}
    \begin{split}
        &\;\;\;u_{IR,n}^{a_1\cdots a_{n+2}}=
        \Tr \phi_{IR}^{(a_1}\cdots\phi_{IR}^{a_{n+2})}
        =M^{-\frac{n+2}{4}}u^{a_1\cdots a_{n+2}}\;,
        % \\
        % &v_{IR,n}^{a_1\cdots a_{n+2}}=\Tr \phi_{IR}^{(a_1}\cdots \phi_{IR}^{a_{n+1}}\psi_{IR}^{a_{n+2})}
        % =M^{-\frac{n}{4}}\epsilon^{a3(a_1}v^{a_2a_3\cdots a_{n+2})}\!_{a},\\
        % w_{n}\equiv&\Tr(f(\phi\phi)^{n+1}+\frac{1}{4}\sum_{k=0}^n(\phi\phi)^{n-k}\psi^a(\phi\phi)^{k}\psi_{a}) \ .
    \end{split}
\end{equation}
where $u^{a_1\cdots a_{n+2}}$ is part of $\C{N}=4$ gravitons
\begin{align}
    u^{m_1\cdots m_{n+2}}=\Tr \phi^{(m_1}\cdots \phi^{m_{n+2})}\;,
\end{align}
Here, $a_k=1,2$ while $m_k=1,2,3$.
In other words, there is no need to take the $M\to 0$ limit in order to obtain UV cohomology.
The UV expression of $v_{IR,n}$ is slightly less trivial, but one can show that 
\begin{align}
    v_{IR,n}^{a_1\cdots a_{n+2}}=\Tr \phi_{IR}^{(a_1}\cdots \phi_{IR}^{a_{n+1}}\psi_{IR}^{a_{n+2})}
    =2M^{-\frac{n}{4}}\epsilon^{b3(a_1}v^{a_2\cdots a_{n+2})}\!\,_{b}\;,
\end{align}
where $v^{a_1\cdots a_{n+2}}\!\,_b$ is part of $\C{N}=4$ gravitons
\begin{align}
    v^{m_1\cdots m_k}\!\,_{m_{k+1}}=\Tr \phi^{(m_1}\cdots \phi^{m_k)}\psi_{m_{k+1}}-(traces)\;,
\end{align}
Here, $(traces)$ means that one has to subtract trace parts such that the contractions of $m_i$ ($i=1,...,k$) with $m_{k+1}$ vanish.

Nontrivial cases start from $w_{IR,n}$.
In terms of the UV letters, the lowest graviton $w_{IR,0}$ takes the form
\begin{equation}
    \begin{split}
        w_{IR,0}&=\Tr \!\,\big(f_{IR}\phi^a_{IR}\phi_{IR\,a}+\frac{1}{4}\psi^a_{IR}\psi_{IR\,a}\big)\\
        &=\frac{4}{M^{3/2}}\Tr((\phi\phi)\psi_{3}\psi_{3})+\frac{2}{M^{1/2}}\Tr(f(\phi\phi)-[\phi^a,\psi_{a}]\psi_{3})-2M^{1/2}\Tr(\psi_{1}\psi_{2}) \ .
    \end{split}
\end{equation}
This is in the general form \eqref{eq: IR op in UV}.
Substituting the relation $(\phi\phi)=-[X,Y]=MZ-Q\psi_{3}$, one can show
\begin{align}
    w_{IR,0}=-\frac{4}{3M^{3/2}}Q\Tr(\psi_{3})^3
    +\frac{2}{M^{1/2}}Q\Tr(f\psi_{3})
    -2M^{1/2}\Tr(Zf+\psi_{1}\psi_{2}) \ .
\end{align}
The last term, of order $M^{1/2}$, is a component of an $\C{N}=4$ graviton operator.
\begin{align}
    w^m=\Tr(\phi^m f+\frac{1}{2}\epsilon^{mnp}\psi_n\psi_p)
\end{align}
Thus, we have shown that the IR SCFT graviton $w_0$ becomes the $\mathcal{N}=4$ SYM graviton \cite{Choi:2023vdm}, up to $Q$-exact terms (not up to $Q_0$-exact).
We expect that the higher gravitons $w_{IR,n\geq 1}$ become part of the higher $\C{N}=4$ gravitons ($m_1=...=m_{n+1}=3$ component)
\begin{align}
    w^{m_1\cdots m_{n+1}}=\Tr\;\!\Big[\phi^{(m_1}\cdots\phi^{m_{n+1})}f
    +\frac{1}{2}\psi_m \sum_{k=1}^{n+1}\phi^{(m_1}\cdots\phi^{m_{k-1}}
    \epsilon^{m_k|mp}\psi_p \phi^{|m_{k+1}}\cdots \phi^{m_{n+1})}
    \Big]\;,
\end{align}
up to $Q$-exact terms.

So far, we have seen that the IR gravitons come from a subset of UV gravitons.
It is natural to question the fate of the rest of the UV gravitons.
At the UV with small mass, one can explicitly see that the gravitons containing $Z$ or $\psi_3$ are lifted. For example,
\begin{align}
    u^{m3}=(1+\frac{\delta^{m}_3}{2})\frac{1}{M}Qv^{m}\!_3\;,\;\;
    v^3\!_a=-\frac{1}{M}\epsilon_{ab}Qw^b\;.
\end{align}
Roughly speaking, graviton operators containing $\psi_3$ become non-closed, and graviton operators containing $Z$ become exact.
Note that when an $\C{N}=4$ cohomology contains $\psi_3$, the mass-deformed $Q$-action effectively becomes $M Q_1$ because it is closed under $Q_0$.

% (\textbf{JC: It would be very nice if we could do similar things with $w_{n\geq 1}$, because the similar analysis could imply that $w_n$s' are non-exact for any $n\geq 1$—which is non-trivial.... Actually, there is no implication for non-exactness because $MZ=Q\psi_3$ for diagonal fields.})

%%%%%%%%%%%%%%%%%%%%%%%%%%%%%%%%%%%%%%%%%
%%%%%%%%%%%%%%%%%%%%%%%%%%%%%%%%%%%%%%%%%
%%%%%%%%%%%%%%%%%%%%%%%%%%%%%%%%%%%%%%%%%
%%%%%%%%%%%%%%%%%%%%%%%%%%%%%%%%%%%%%%%%%
\subsection{Coulomb non-gravitons}
We now specialize to the $SU(2)$ gauge group. To express the UV and IR fields in vector notation, we note that
\begin{equation}
    \begin{aligned}
        \phi^{a}_{IR}=\frac{i}{2}\phi^a_{IR}\cdot\vec{\sigma}\;,\;\;\;
        \psi_{IR\,a}=
        &i\psi_{IR\,a}\cdot\vec{\sigma}\;,\;\;\;
        f_{IR}=
        -if_{IR}\cdot\vec{\sigma},\\
        \phi^{m}=\frac{i}{2}\phi^m\cdot\vec{\sigma}\;,\;\;\;
        \psi_{m}=&\frac{i}{2}\psi_{m}\cdot\vec{\sigma}\;,\;\;\;
        f=\frac{i}{2}f\cdot\vec{\sigma},
    \end{aligned}
\end{equation}
The $Q$-actions on the UV letters are then
\begin{equation}
    \begin{aligned}
        Q\psi_{m}=-\frac{1}{2}\epsilon_{mnp}\phi^n\times\phi^p
        +M\delta_{m,3}\phi^3\;,\;\;\;
        Qf=-\phi^m\times\psi_m\;\;,
    \end{aligned}
\end{equation}
and on the IR letters,
\begin{align}
    Q\psi_{IR\,a}=\phi_{IR\,a}\times(\phi^1_{IR}\times\phi^2_{IR})\;,\;\;\;
    Qf=-\phi^a_{IR}\times\psi_{IR\,a}\;,
\end{align}
This is indeed the convention adopted throughout this paper. 

The transformation rules relating IR and UV letters are then given by
\begin{align}
    \phi_{IR}^a=M^{-1/4}\phi^a,\quad
    \psi_{IR\,a}=M^{1/4}(\psi_a-\frac{\phi_a\times\psi_3}{M}),\quad
    f_{IR}=f+\frac{\psi_3\times\psi_3}{2M}\ .
\end{align}
Substituting these into $O_7$ and expanding in powers of $M$, the $M^{-7/4}$ terms cancel, and
\begin{align}
    M^{3/4}O_7^{abc}=
    \phi^{(a}\cdot\phi^b
    (\phi^{c)}\cdot f+\psi^{c)}\cdot\psi_3)
    +\phi^{(a}\cdot\psi_3\phi^{b}\cdot\psi^{c)}
    -\frac{M}{2}\phi^{(a}\cdot\psi^{b}\times\psi^{c)}.
\end{align}
Now, in terms of the $SU(2)$ expressions of UV gravitons,
\begin{align}
    u^{mn}=\phi^m\cdot\phi^n\;,\;\;
    v^{m}\!_n=\phi^m\cdot\psi_n
    -\frac{\delta^m_n}{3}\phi^p\cdot\psi_p\;,\;\;
    w^{m}=\phi^m\cdot f+\frac{1}{2}\epsilon^{mnp}
    \psi_n\cdot\psi_p\;,
\end{align}
the fortuitous operator $O_7$ becomes
\begin{align}\label{eqn: O7 using N=4 garv}
    M^{3/4}O_7^{abc}
    =u^{(ab}w^{c)}+v^{(a}\!_{3} v^{bc)}
    -\frac{M}{2}\phi^{(a}\cdot\psi^b\times\psi^{c)}\;.
\end{align}
where $v^{ab}\equiv v^{(a}\!_{m}\epsilon^{b)m3}$.
Similarly, for $O_9$, all contributions of order $M^{-9/4}$ and $M^{-5/4}$ cancel, and the remainder is
\begin{equation}\label{eqn: O9 using N=4 garv}
\begin{aligned}
    M^{1/4}O^{abc}_9
    &=\phi^{(a}\cdot\psi^b(\phi^{c)}\cdot f+\psi^{c)}\cdot\psi_3)
    -\frac{M}{6}\psi^{(a}\cdot\psi^b\times\psi^{c)}\\
    &=v^{(ab}w^{c)}
    -\frac{M}{6}\psi^{(a}\cdot\psi^b\times\psi^{c)} \ ,
\end{aligned}
\end{equation}
Also for $O_{12}$, terms of order $M^{-3}$ and $M^{-2}$ vanish, and the remaining ones are:
\begin{equation}\label{eqn: O12 using N=4 garv}
    \begin{split}
        M O_{12}^{abcd}&=(f\cdot\phi^{(a})(f\cdot\phi^{b})(\phi^c\cdot\phi^{d)})-\frac{1}{2}(\phi^{(a}\cdot\psi_3\times\psi_3)(\phi^{b}\cdot\psi^c\times\psi^{d)})\\
        &\qquad+2(f\cdot\phi^{(a})(\phi^b\cdot\phi^c)(\psi^{d)}\cdot\psi_3)+2(f\cdot\phi^{(a})(\phi^b\cdot\psi_3)(\phi^{c}\cdot\psi^{d)})\\
        &\qquad-M\phi^{(a}\cdot f\phi^b\cdot\psi^c\times\psi^{d)}\\
        &=u^{(ab}w^cw^{d)}
    -2w^{(a}v^{bc}v^{d)}\!_3
    -M\phi^{(a}\cdot f\phi^b\cdot\psi^c\times\psi^{d)} \ .
    \end{split}
\end{equation}
For $O_{14}$, the terms of order $M^{-5/2}$ and $M^{-3/2}$ cancel out, yielding
\begin{equation}\label{eqn: O14 using N=4 garv}
    M^{1/2}O_{14}^{abcd}=w^{(a}w^{b}v^{cd)}-\frac{M}{3}f\cdot\phi^{(a}\psi^{b}\cdot\psi^c\times\psi^{d)}\;.
\end{equation}

\subsection{Non-Coulomb non-gravitons}
Finally, we would like to analyze the non-Coulomb, non-graviton operator $O_{16}$ \eqref{eq: O16}. In this case, the terms of order $M^{-4}$ and $M^{-3}$ vanish, and the remaining terms take the following form
\begin{equation}
    O_{16}=M^{-2} O_{t^{28}}+M^{-1} O_{t^{26}}+ O_{t^{24}}
    +M O_{t^{22}}
\end{equation}
The explicit expressions of $O_{t^{22}},O_{t^{24}},O_{t^{26}},O_{t^{28}}$ are in Appendix \ref{appendix: O16UV}, which are quite lengthy.

As mentioned in the introduction, $O_{16}$ has IR R-charge $-3$ and $j=5/2$, and it contributes to the index at $-t^{24}$.
Since $M$ has UV R-charge $-2/3$, it carries $t^{3(2j-r^{UV})}=t^2$, which means that the UV operators have charges such that $t^{3(2j-r^{UV})}=t^{28},t^{26},t^{24},t^{22}$, all being fermionic.
The charges of operators are as follows.
\begin{table}[H]
    \centering
    \begin{tabular}{c||c|c|c|c||c}
      & $O_{t^{28}}$ & $O_{t^{26}}$ & $O_{t^{24}}$ & $O_{t^{22}}$ & $O_{16}$ \\ \hline
      $r^{UV}$ & $-\frac{13}{3}$ & $-\frac{11}{3}$ & $-3$ & $-\frac{7}{3}$ & $-3$ \\ \hline
      $j$ & $\frac{5}{2}$ & $\frac{5}{2}$ & $\frac{5}{2}$ & $\frac{5}{2}$ & $\frac{5}{2}$ \\ \hline
      $L^{UV}$ & $9$ & $8$ & $7$ & $6$ &
    \end{tabular}
    \caption{Charges of UV operators and $O_{16}$. $L^{UV}$ stands for the number of UV letters. $O_{16}$ has ill-defined $L^{UV}$. In fact, since the IR $Q$-action increases the number of IR letters either by two or one, the number of IR letters of $O_{16}$ is also ill-defined.}
    \label{tab: charges of O_UV}
\end{table}\vspace{-0.3cm}
% \noindent As a first guess, 

We want to show that this IR cohomology $O_{16}$ becomes the lightest non-graviton cohomology of $\C{N}=4$ SYM which was named as `$O_0$' in \cite{Choi:2023znd}, up to $Q$-exactness.
Rather than proving via a brute-force method, we argue slightly indirectly as follows.

% (\textbf{JC: $-q^{16}=-t^{24}$ and $M$ has $t^{2}$ hence $M^2 O_{16}$ is at $t^{28}$. Could it be a super-descendant of `$O_0$' of $SU(2)$ MSYM? i.e. $Q_+^1 Q_+^2 O_0$ to be singlet under $SU(2)_F$.
% This can't be true due to the eqn(3.21) of \cite{Choi:2023znd} — explicit expression of $Q^1_+Q^2_+ O_0'$. The expression contains a term of the form $\psi^7$ and this can't be cancelled by adding a $Q$-exact term. Since our operator $O_{16}$ doesn't contain $\psi$-only or $\psi,f$-only term, UV expression can never produce $\psi^7$ term.}) :)

% (\textbf{JC: remaining possibilities are (1) $M^{-1}$ part of $O_{16}$ is cohomologous to $Q^3_+O_0$ of MSYM, (2) $M^0$ part of $O_{16}$ is cohomologous to $O_0$ (3) $O_{16}$ somehow becomes graviton albeit it is non-eigenvalue. The third possibility seems unlikely because it should vanish order by order in $M$ when UV fields are diagonal which implies IR fields being diagonal.})
% :)

% (\textbf{JC: The leading order in small $M$ expansion should be annihilated by $M^0$ part of $Q$ i.e. $Q$ of MSYM. Therefore, $M^{-2}$ part of $O_{16}$ must be $Q$-exact in MSYM.})

% (\textbf{JC: $O_{16}$ is at $-t^{24}$ with $r=-3, j=5/2$ and $O_0$ of MSYM is also at $-t^{24}$ with $r=-3,j=5/2$. On the other hand, since $r[Q_+^m]=1/3$, $Q_{+}^3O_0$ is at $+t^{26}$ with $r=-3+1/3,j=6/2$ and $Q_+^1Q_+^2O_0$ is at $-t^{28}$ with $r=-3+2/3,j=7/2$. Therefore, $O_{t^{26}}$ and $O_{t^{28}}$—both fermionic and $j=5/2$— can't be related to descendants of $O_0$ of MSYM.})

From the $Q$-closedness of $O_{16}$,
\begin{align}
    (Q_0+MQ_1)O_{16}=&
    M^{-2}Q_0O_{t^{28}}+M^{-1} (Q_1O_{t^{28}}+ Q_0O_{t^{26}})\nonumber\\
    +& (Q_1O_{t^{26}}+Q_0O_{t^{24}})
    +M (Q_1O_{t^{24}}+Q_0O_{t^{22}})+M^2Q_1 O_{t^{22}}=0
    \;.
\end{align}
we see that $Q_0 O_{t^{28}}=0$. 
However, we know from the literature \cite{Chang:2022mjp,Choi:2023znd} that there is no non-trivial $Q_0$-cohomology with $9$ letters with charges $r=-\frac{13}{3}$ and $j=\frac{5}{2}$ in $SU(2)$ $\C{N}=4$ SYM.
So $O_{t^{28}}$ is $Q_0$-exact, 
$$O_{t^{28}}=Q_0 \tilde{O}_{t^{28}},$$
and it can be naturally mapped to a $Q$-exact operator, just by acting $Q$ on $\tilde{O}_{t^{28}}$.
However, $\tilde{O}_{t^{28}}$ is ambiguous because one can add any $Q_0$-closed term to it, which might not be $Q_1$-closed. Therefore, the map to $Q$-exact operator is also ambiguous in this sense.
In any case, if we subtract $M^{-2}Q\tilde{O}_{t^{28}}$ from $O_{16}$,
$$O_{16}-M^{-2}Q\tilde{O}_{t^{28}}=M^{-1}(O_{t^{26}}-Q_1\tilde{O}_{t^{28}})+ O_{t^{24}}
+MO_{t^{22}}\;,$$
% The $Q$-action on this is
% \begin{equation}
% \begin{aligned}
%     0=M(Q_0O_{t^{26}}-Q_0Q_1\tilde{O}_{t^{28}})
%     +M^2(Q_1O_{t^{26}}+Q_0O_{t^{24}})
%     +M^3 (Q_1O_{t^{24}}+Q_0O_{t^{22}})\;.
% \end{aligned}
% \end{equation}
By the same reasoning, we see that $(O_{t^{26}}-Q_1\tilde{O}_{t^{28}})$ is $Q_0$-closed. This fact is independent of the ambiguity of $\tilde{O}_{t^{28}}$.
Since it is known that there is no non-graviton $Q_0$-cohomology in the charge sector ($r=-\frac{11}{3}$, $j=\frac{5}{2}$) to which $O_{t^{26}}$ belongs, it is either a UV graviton or $Q_0$-exact.
Since the number of letters is $8$, it can be graviton, as the gravitons exist for even number of letters in $SU(2)$ $\C{N}=4$ SYM. It can be either $SU(2)_F$ singlet combination of $u v w^2$ or $v^3w$.
However, from the explicit expression of $O_{t^{26}}$ \eqref{eq: Ot26}, we see that it is non-Coulomb. This implies that the possibility of $(O_{t^{26}}-Q_1\tilde{O}_{t^{28}})$ being a graviton is only by the $Q_1\tilde{O}_{t^{28}}$ part.
Here, we note that the ambiguity of $\tilde{O}_{t^{28}}$ is either $Q_0$-exact or gravitons, because there is no non-graviton $Q_0$-cohomology at $r^{UV}=-\frac{10}{3}$, $L^{UV}=8$, and $j=3$ (charges of $\tilde{O}_{t^{28}}$). But the $Q_0$-exact operator is still $Q_0$-exact after the action of $Q_1$ because $\{Q_0,Q_1\}=0$. Therefore, any $Q_1$-exact graviton at $r=-\frac{11}{3},j=\frac{5}{2},L=8$ is $Q_1$-action on another graviton ($\sim v^2w^2$), and it can be eliminated by the choice of $\tilde{O}_{t^{28}}$.

Thus, $O_{t^{26}}-Q_1\tilde{O}_{t^{28}}$ can be chosen to be $Q_0$-exact:
\begin{align}\label{eq:O0with mass}
    O_{t^{26}}-Q_1\tilde{O}_{t^{28}}=Q_0\tilde{O}_{t^{26}}\;,
\end{align}
We can again map it to a $Q$-exact operator by acting $Q$ on $\tilde{O}_{t^{26}}$.
Then,
$$O_{16}-M^{-2}Q\tilde{O}_{t^{28}}-M^{-1}Q\tilde{O}_{t^{26}}
=(O_{t^{24}}-Q_1\tilde{O}_{t^{26}})+M O_{t^{22}}\;,$$
From this, $O_{t^{24}}-Q_1\tilde{O}_{t^{26}}$ is $Q_0$-closed.
It has charges $r=-3,j=\frac{5}{2},L=7$. There is one non-graviton $Q_0$-cohomology `$O_0$', and there is no graviton because $L$ is odd. So it is either $O_0$ or $Q_0$-exact.

Let us prove that it is not $Q_0$-exact as follows.
Suppose that $O_{t^{24}}-Q_1\tilde{O}_{t^{26}}$ is $Q_0$-exact; $Q_0\tilde{O}_{t^{24}}$.
Then, 
$$O_{16}-M^{-2}Q\tilde{O}_{t^{28}}-M^{-1}Q\tilde{O}_{t^{26}}-Q\tilde{O}_{t^{24}}
=M(O_{t^{22}}-Q_1\tilde{O}_{t^{24}})\;,$$
$(O_{t^{22}}-Q_1\tilde{O}_{t^{24}})$ is $Q_0$-closed, $Q_1$-closed, and $L=6$.
Since $O_{t^{22}}$ is non-Coulomb from the explicit expression, the possibility of being a graviton can be eliminated by a suitable choice of $\tilde{O}_{t^{24}}$, following the same logic used above.
This means that there is a choice of $\tilde{O}_{t^{24}}$ such that 
$$(O_{t^{22}}-Q_1\tilde{O}_{t^{24}})=Q_0\tilde{O}_{t^{22}}\;,$$
We can again subtract $Q\tilde{O}_{t^{22}}$, and we get $-M^2 Q_1\tilde{O}_{t^{22}}$.
By repeating the same argument, it is also $Q_0$-exact; $-Q_1\tilde{O}_{t^{22}}=Q_0\tilde{O}_{t^{20}}$.
Further subtracting $Q\tilde{O}_{t^{20}}$, we obtain $-M^3Q_1\tilde{O}_{t^{20}}$, which is $Q_0$-closed, and $L=4$. Again it should be $Q_0$-exact; $-Q_1\tilde{O}_{t^{20}}=Q_0\tilde{O}_{t^{18}}$, where $\tilde{O}_{t^{18}}$ has $L=3$.
But there is no gauge invariant operator at $t^{18}$ and $L=3$. The only candidate is $f\cdot f\times f=0$, meaning $Q_1\tilde{O}_{t^{20}}=0$.
From this, we conclude that $O_{16}$ is $Q$-exact, hence a contradiction.

Therefore, we have shown that $O_{t^{24}}-Q_1\tilde{O}_{t^{26}}$ is the lightest fortuitous operator `$O_0$' in $SU(2)$ $\C{N}=4$ SYM, up to $Q_0$-exactness.
Also, the right hand side of \eqref{eq:O0with mass} can be regarded as the mass-deformed version of $O_0$ of $\C{N}=4$ SYM.

\section{Discussion}\label{section: discussion}
In this work, we analyzed the $Q$-cohomology of the $\C{N}=1$ SCFT obtained from the mass deformation of $\C{N}=4$ SYM, focusing on the BMN sector and constructing explicit representatives of both graviton and non-graviton operators.
We found that there is a useful change of variables that makes the cohomology problem much more tractable.
A notable feature of the deformed theory is that, unlike in $SU(2)$ and $SU(3)$ $\C{N}=4$ SYM, the distinction between Coulomb and non-Coulomb cohomologies no longer aligns with the graviton/non-graviton classification. 
We encountered all four types of cohomologies: Coulomb graviton, non-Coulomb graviton, Coulomb non-graviton, and non-Coulomb non-graviton.
For the examples that we have found in $SU(2)$, we have shown that Coulomb non-gravitons are mass-deformed versions of $\C{N}=4$ gravitons, even though they are fortuitous.
We have also shown that the non-Coulomb non-graviton $O_{16}$ reduces to the lightest non-graviton of $SU(2)$ $\C{N}=4$ SYM.
This indicates that the fortuity can emerge due to a relevant deformation. But roughly speaking, such emergent fortuitous cohomologies still behave like gravitons in the sense that they are of Coulomb type.

Our analysis was performed entirely at the level of classical cohomology.
It would be interesting to investigate how robust our results are (e.g. Coulomb graviton index, cohomology classes). Since the theory is intrinsically strongly coupled, the effect of quantum corrections to the cohomologies can be more severe than those in $\C{N}=4$ SYM \cite{Choi:2025bhi}.

% We have found non-graviton cohomologies, but there is no known type $\rm IIB$ black hole solution in the asymptotic Pilch-Warner background.

We speculate that the existence of non-Coulomb graviton cohomologies is related to the presence of a conical singularity at the origin of the Coulomb branch moduli space.
As mentioned in Section \ref{sub:coulomb noncoulomb}, a probe D3-brane placed in $AdS_5$ feels the moduli space of dimension 4, with a conical singularity at the origin.
Due to this singularity, some excitations of the probe D3-brane might be localized near the origin of moduli space, at which off-diagonal elements of the fields are light. We suspect that the non-Coulomb gravitons (which could be thought of as an excitation of a small D3-brane) like $w_n$ might be related to such excitations.

% We have not attempted to provide a definitive explanation here, but one may speculate that the very different behavior of eigenvalues in the UV and IR is related holographically to the distinct probe-brane geometries. For example, the presence of non-eigenvalue gravitons in the IR SCFT might be connected to the fact that the moduli space of D3-brane configurations in the Pilch–Warner background develops a conical singularity at the origin. At this singular point, where the IR SCFT with $SU(2)$ gauge group is realized, new massless degrees of freedom emerge. These modes activate additional off-diagonal components of the massless scalars, so that the gravitons constructed from them can no longer be described purely in terms of eigenvalues. Importantly, these new degrees of freedom do not originate from the orbifold singularities associated with the $S_N$ quotient. It may be possible to analyze such a singularity using an effective theory localized near the singular point to better understand the underlying mechanism.

Another possible explanation, though even more tentative, for non-Coulomb gravitons is the Myers effect \cite{Myers:1999ps, Polchinski:2000uf}; D3-brane polarization induced by the nontrivial two-form fields in the background. In this picture, single-gravitons are considered as single D3-brane states, and their positions no longer commute with each other, requiring off-diagonal elements of the adjoint fields. 
% Of course, in our case the IR SCFT has a continuous moduli space, unlike the setups considered in \cite{Polchinski:2000uf}, which involved a discrete and isolated set of vacua. 
% More precisely, they studied the so-called $\C{N}=1^*$ theory, obtained by turning on finite mass parameters for all chiral fields of $\C{N}=4$ SYM. This theory also flows to an IR SCFT, but with confining or Higgs vacua.
It would be very interesting to perform probe-brane computations in the Pilch–Warner background, which has nontrivial flux, following \cite{Polchinski:2000uf}.

Whether every Coulomb non-graviton originates from a UV graviton is unclear, although all the examples we have found are consistent with this.
Since UV letters are diagonal when IR fields are diagonal, an IR Coulomb cohomology becomes a UV Coulomb {operator}. However, since the $Q$-action of mass-deformed $\C{N}=4$ is not necessarily a commutator, the concept of Coulomb \emph{cohomology} is not well-defined in the UV coordinates.
It is also unclear whether non-Coulomb non-gravitons must always originate from UV non-gravitons, and it would be interesting to prove or disprove this relation.

Our analysis of fortuitous cohomologies was carried out explicitly for gauge group $SU(2)$. While this already reveals nontrivial features of the deformation, it would be interesting to extend the construction to higher-rank gauge groups.
It would also be interesting to understand whether the cohomological integrating-out procedure (of massive fields) can be generalized to other theories with less supersymmetry, e.g., the Klebanov-Witten theory \cite{Klebanov:1998hh} as a relevant deformation of the $\C{N}=2$ orbifold theory \cite{conifold}.

\begin{acknowledgments}
We thank Minseok Cho, Jaewon Song, and Seok Kim for helpful discussions. We are particularly grateful to Jaewon Song for originally suggesting this project, and to Seok Kim for suggesting that we derive the $Q$-action directly from the UV mass-deformed Lagrangian.
The work of J.C was supported by a KIAS Individual Grant PG106601 at the Korea Institute for Advanced Study.
The work of S.K. is supported by the National Research Foundation of Korea (NRF) Grant RS-2024-00412099 and RS-2023-00208602 at KAIST.

\end{acknowledgments}

\appendix

%%%%%%%%%%%%%%%%%%%%%%%%%%%%%%%%%%%%%%%%%
%%%%%%%%%%%%%%%%%%%%%%%%%%%%%%%%%%%%%%%%%
%%%%%%%%%%%%%%%%%%%%%%%%%%%%%%%%%%%%%%%%%
%%%%%%%%%%%%%%%%%%%%%%%%%%%%%%%%%%%%%%%%%
%%%%%%%%%%%%%%%%%%%%%%%%%%%%%%%%%%%%%%%%%
%%%%%%%%%%%%%%%%%%%%%%%%%%%%%%%%%%%%%%%%%

% \section{Hodge theory argument}\label{apndx:hodge}
% Any local operator $O$ in SCFT can be expanded in eigenstates of $\Delta\equiv 2\{Q,Q^\dag\}$ where the dagger is in radial quantization $i.e.$ $Q^\dag=S$. Therefore, $$O=O_{BPS}+\sum_i O^{(i)}_{non-BPS}.$$
% Let's say the eigenvalues of non-BPS operators of $\Delta$ are $\delta_i$. Then, $$O=O_{BPS}+\frac{\Delta}{2} O_{non-BPS},$$ where $O_{non-BPS}=\sum_i \frac{2}{\delta_i}O_{non-BPS}^{(i)}$.
% If $O$ is $Q$-closed, 
% $$||Q^\dag Q O_{non-BPS}||^2=
% \langle O_{non-BPS}|Q^\dag Q \big(|O\rangle-|O_{BPS}\rangle-Q Q^{\dag}|O_{non-BPS}\rangle\big)
% =0.$$
% Therefore,
% $$O=O_{BPS}+Q(2Q^{\dag}O_{non-BPS}).$$
% In other words, there is a unique BPS operator (including the trivial case) in any $Q$-cohomology class.

%%%%%%%%%%%%%%%%%%%%%%%%%%%%%%%%%%%%%%%%%
%%%%%%%%%%%%%%%%%%%%%%%%%%%%%%%%%%%%%%%%%
%%%%%%%%%%%%%%%%%%%%%%%%%%%%%%%%%%%%%%%%%
%%%%%%%%%%%%%%%%%%%%%%%%%%%%%%%%%%%%%%%%%
%%%%%%%%%%%%%%%%%%%%%%%%%%%%%%%%%%%%%%%%%
%%%%%%%%%%%%%%%%%%%%%%%%%%%%%%%%%%%%%%%%%
\section{$w_n$ for $n\geq 1$ in $SU(2)$}\label{minimal check}
In this appendix, we explicitly demonstrate that the single-trace graviton operators $w_n$ for $n\geq 1$ become $Q$-exact when the gauge group is $SU(2)$. 
% Since they vanish when the fields are diagonal (except when $n = 0$), there is no guarantee that these are not $Q$-exact.
% we expect them to be $Q$-exact in $SU(2)$ to be consistent with the eigenvalue counting method. 
To proceed, we divide the tower of operators into two cases: odd $n = 2m + 1$ and even $n = 2m$.
\begin{equation}\label{eqn: single-trace graviton vanishing SU(2) 2}
    \begin{split}
        w_{2m+1}=&\Tr(f(\phi\phi)^{2m+2}+\frac{1}{4}\sum_{k=0}^{2m+1}(\phi\phi)^{2m+1-k}\psi^a(\phi\phi)^k\psi_{a})\qquad (m\geq0) \ ,\\
        w_{2m}=&\Tr(f(\phi\phi)^{2m+1}+\frac{1}{4}\sum_{k=0}^{2m}(\phi\phi)^{2m-k}\psi^a(\phi\phi)^k\psi_{a})\qquad (m\geq0) \ .
    \end{split}
\end{equation}
We begin by showing that the first line vanishes when the gauge group is $SU(2)$. In terms of vector notation,
\begin{equation}
    \Tr(f(\phi\phi)^{2m+2})=-i\left(\frac{i}{2}\right)^{2m+2}(\phi^1\times\phi^2)_{i_1}\cdots (\phi^1\times\phi^2)_{i_{2m+2}}f_{2m+3}\Tr(\sigma_{(i_1}\cdots\sigma_{i_{2m+2})}\sigma_{i_{2m+3}})
\end{equation}
since $f=-if\cdot\vec{\sigma}$ and
\begin{equation}
    (\phi\phi)=[\phi^2,\phi^1]=\left(\frac{i}{2}\right)^2(\phi^2)_i(\phi^1)_j[\sigma_i,\sigma_j]=-\frac{i}{2}\epsilon_{ijk}(\phi^2)_i(\phi^1)_j\sigma_k=\frac{i}{2}(\phi^1\times\phi^2)\cdot\vec{\sigma}
\end{equation}
Now, using the identities
\begin{align}
    \sigma_{(i_1}\cdots \sigma_{i_{2k})}
    &=\delta_{(i_1i_2}\cdots\delta_{i_{2k-1}i_{2k})},\\
    \sigma_{(i_1}\cdots \sigma_{i_{2k+1})}
    &=\delta_{(i_1i_2}\cdots\delta_{i_{2k-1}i_{2k}}\sigma_{i_{2k+1})}\,
\end{align}
we find
\begin{equation}
    \begin{split}
        &\Tr(f(\phi\phi)^{2m+2})\\
        &=i\left(\frac{i}{2}\right)^{2m+2}(\phi^1\times\phi^2)_{i_1}\cdots (\phi^1\times\phi^2)_{i_{2m+2}}f_{i_{2m+3}}\delta_{(i_1i_2}\cdots\delta_{i_{2k-1}i_{2k+2})}\Tr(\sigma_{i_{2k+3}})\\
        &=0
    \end{split}
\end{equation}
The second term, on the other hand, can be written as
\begin{align}
    &\Tr (\phi\phi)^{2m+1-k}\psi_+^a(\phi\phi)^k\psi_{a}
    \nonumber\\
    &\qquad=\left(\frac{i}{2}\right)^{2m+1}(\phi^1\times\phi^2)_{i_1}\cdots(\phi^1\times\phi^2)_{i_{2m+1}}\psi^a_{i_{2m+2}}\psi_{a,i_{2m+3}}\\
    &\qquad\times\Tr(\sigma_{(i_1}\cdots \sigma_{i_{2m+1-k}| }\sigma_{(i_{2m+2}|}\sigma_{|i_{2m+1-k+1}}\cdots\sigma_{i_{2m+1})}\sigma_{|i_{2m+3})})\nonumber \ ,
\end{align}
where
\begin{align}
    &\Tr(\sigma_{(i_1}\cdots \sigma_{i_{2m+1-k}| }\sigma_{(i_{2m+2}|}\sigma_{|i_{2m+1-k+1}}\cdots\sigma_{i_{2m+1})}\sigma_{|i_{2m+3})})\\
    &=\begin{cases}
    \Tr(\delta_{(i_1i_2}\cdots \delta_{i_{2m-k}i_{2m+1-k}|} \sigma_{(i_{2m+2}|}
    \delta_{|i_{2m+2-k}i_{2m+3-k}}\cdots
    \delta_{i_{2m-1}i_{2m}}\sigma_{i_{2m+1})}
    \sigma_{|i_{2m+3})})
    \quad \text{($k$ odd)}\\
    \Tr(\delta_{(i_1i_2}\cdots 
    \delta_{i_{2m-1-k}i_{2m-k}}\sigma_{i_{2m+1-k}|} 
    \sigma_{(i_{2m+2}|}
    \delta_{|i_{2m+2-k}i_{2m+3-k}}\cdots
    \delta_{i_{2m}i_{2m+1})}
    \sigma_{|i_{2m+3})})
    \quad \text{($k$ even)}
    \end{cases}\nonumber
\end{align}
In both cases, the trace vanishes because it eventually reduces to a Levi-Civita symbol with some symmetrized indices, which is identically zero.

Next, we demonstrate that the second line of \eqref{eqn: single-trace graviton vanishing SU(2) 2} is $Q$-exact in $SU(2)$. The expression can be rewritten in the form
\begin{equation}
    w_{2m}=\Tr(f(\phi\phi)^{2m+1}+\frac{1}{2}\sum_{k=0}^{m-1}(\phi\phi)^{2m-k}\psi^a(\phi\phi)^k\psi_{a}+\frac{1}{4}(\phi\phi)^m\psi^a(\phi\phi)^m\psi_{a})\ ,
\end{equation}
where the last corresponds to the $k = m$ contribution, separated explicitly from the summation. Note that when $m = 0$, the summation is absent.

For the first term
\begin{equation}
    \begin{split}
        &\Tr(f(\phi\phi)^{2m+1})\\
        &=-i\left(\frac{i}{2}\right)^{2m+1}(\phi^1\times\phi^2)_{i_1}\cdots(\phi^1\times\phi^2)_{i_{2m+1}}f_{i_{2m+2}}\Tr(\delta_{(i_1i_2}\cdots\delta_{i_{2m+2}i_{2m+3})})\\
        &=-2i\left(\frac{i}{2}\right)^{2m+1}(\phi^1\times\phi^2)^{2m}\phi^1\times\phi^2\cdot f \ .
    \end{split}
\end{equation}
For the second term, for even $k$, we have
\begin{align}
    &\frac{1}{2}\Tr((\phi\phi)^{2m-k}\psi^a(\phi\phi)^k\psi_{a})\\
    &=\frac{1}{2}\left(\frac{i}{2}\right)^{2m}(\phi^1\times\phi^2)_{i_1}\cdots(\phi^1\times\phi^2)_{i_{2m}}\psi^a_{i_{2m+1}}\psi_{a,i_{2m+2}}\Tr(\delta_{(i_1 i_2}\cdots\delta_{i_{2m-1}i_{2m})}\delta_{(i_{2m+1}i_{2m+2})})\nonumber\\
    &=\left(\frac{i}{2}\right)^{2m}(\phi^1\times\phi^2)^{2m}\psi^a\cdot\psi_a \nonumber \ ,
\end{align}
while for odd $k$,
\begin{equation}
    \begin{split}
        &\frac{1}{2}\Tr(\phi\phi)^{2m-k}\psi^a(\phi\phi)^k\psi_{a})\\
        &=\frac{1}{2}\left(\frac{i}{2}\right)^{2m}(\phi^1\times\phi^2)^{2m-2}(\phi^1\times\phi^2)_{i_{2m-k}}\psi^a_{i_{2m+1}}(\phi^1\times\phi^2)_{i_{2m-k+1}}\psi_{a,i_{2m+2}}\\        
        &\qquad \times\Tr(\sigma_{(i_{2m-k}|}\sigma_{(i_{2m+1}|}\sigma_{|i_{2m-k+1})}\sigma_{|i_{2m+2})})\\
        &=\left(\frac{i}{2}\right)^{2m}(\phi^1\times\phi^2)^{2m-2}\big[2(\phi^1\times\phi^2\cdot\psi^a)(\phi^1\times\phi^2\cdot\psi_a)-(\phi^1\times\phi^2)^2\psi^a\cdot\psi_a\big] \ ,
    \end{split}
\end{equation}
where we have used $\Tr(\sigma_i\sigma_j\sigma_k\sigma_l)=2(\delta_{ij}\delta_{kl}-\delta_{il}\delta_{jl}+\delta_{il}\delta_{jk})$. Similarly, for the last term with even $m$:
\begin{align}
    &\frac{1}{4}\Tr((\phi\phi)^m\psi^a(\phi\phi)^m\psi_{a})\\
    &=\begin{cases}
        \frac{1}{2}\left(\frac{i}{2}\right)^{2m}(\phi^1\times\phi^2)^{2m}\psi^a\cdot\psi_a &(\text{even $m$})\\
        \frac{1}{2}\left(\frac{i}{2}\right)^{2m}(\phi^1\times\phi^2)^{2m-2}\big(2(\phi^1\times\phi^2\cdot\psi^a)(\phi^1\times\phi^2\cdot\psi_a)-(\phi^1\times\phi^2)^2\psi^a\cdot\psi_a\big) \;\; &(\text{odd $m$})
    \end{cases}\nonumber
\end{align}
Altogether, we get
\begin{align}
    w_{2m}&=\left(\frac{i}{2}\right)^{2m}(\phi^1\times\phi^2)^{2m-2}\\
    &\times\bigg[(\phi^1\times\phi^2)^2\big(\phi^1\times\phi^2\cdot f-\frac{1}{2}\psi^a\cdot\psi_a\big)+\frac{m}{4}(\phi^1\times\phi^2\cdot\psi^a)(\phi^1\times\phi^2\cdot\psi_a)\bigg]\nonumber \ .
\end{align}
Now using the $Q$-action
\begin{equation}
    Q\psi_a=\phi_a\times(\phi^1\times\phi^2),\qquad Qf=\psi_a\times\phi^a \ ,
\end{equation}
one can easily check that, using $A\cdot(B\times C)=B\cdot(C\times A)=C\cdot(A\times B)$,
\begin{equation}
    Q(\phi^a\cdot\psi_a)=\phi^a\cdot (\phi_a\times(\phi^1\times\phi^2))=-2(\phi^1\times\phi^2)^2\ ,
\end{equation}
and
\begin{equation}
    \begin{split}
        Q(\phi^a\cdot f)
        =\phi^a\cdot\psi_b\times\phi^b=\epsilon^{ba}\psi_b\cdot\phi^1\times\phi^2=-\phi^1\times\phi^2\cdot\psi^a \ ,
    \end{split}
\end{equation}
and therefore $w_{2m}$ is $Q$-exact. For $m=0$, $w_0 =w= \phi^1 \times \phi^2 \cdot f -\frac{1}{2} \psi^a \cdot \psi_a$ is $Q$-closed but not $Q$-exact, as the $Q$-action always increases the number of scalar fields. As a result, $w_{n}$ is $Q$-exact in $SU(2)$ for $n \geq 1$.

\section{UV expression of $O_{16}$}\label{appendix: O16UV}
\begin{equation}
    O_{16}=M^{-2} O_{t^{28}}+M^{-1} O_{t^{26}}+ O_{t^{24}}
    +M O_{t^{22}}
\end{equation}

At order $M^1$, we have
\begin{equation}
\begin{aligned}
    O_{t^{22}}=\frac{1}{3}(\psi^a\cdot\psi^b\times\psi^c)(\psi_a\cdot\psi_b\times\phi_c)\;.
\end{aligned}
\end{equation}
which is annihilated by $Q_1$, as it should be.

At order $M^0$ ($O_{t^{24}}$):
\begin{equation}
    \begin{split}
        (f\cdot\phi^a)(\psi_a\cdot\phi^b\times\phi_b)(\psi^c\cdot\psi_c)\\
        2(f\cdot\psi^a)(\phi_a\cdot\psi^b)(\psi_b\cdot\phi^c\times\phi_c)\\
        -\frac{2}{3}(\psi^a\cdot\psi_3)(\phi^b\cdot\psi^c)(\psi_a\cdot\psi_b\times\phi_c)\\
        +\frac{2}{3}(\psi^a\cdot\psi_3)(\phi^b\cdot\phi^c)(\psi_a\cdot\psi_b\times\psi_c)
    \end{split}
\end{equation}

Order $M^{-1}$ ($O_{t^{26}}$):
\begin{equation}\label{eq: Ot26}
    \begin{split}
        -2(f\cdot\phi^a)(\psi_a\cdot\phi^b\times\phi_b)(\psi^c\cdot\phi_c\times\psi_3)\\
        -2(f\cdot\psi^a)(\phi_a\cdot\phi^b\times\psi_3)(\psi_b\cdot\phi^c\times\phi_c)\\
        -2(f\cdot\phi^a\times\psi_3)(\phi_a\cdot\psi^b)(\psi_b\cdot\phi^c\times\phi_c)\\
        (f\cdot\phi^a\times\phi_a)(f\cdot\phi^b)(\psi_b\cdot\phi^c\times\phi_c)\\
        -\frac{1}{3}(\phi^a\cdot\psi^b)(\phi^c\cdot\psi_3\times\psi_3)(\psi_a\cdot\phi_b\times\psi_c)\\
        -\frac{1}{3}(\psi^a\cdot\psi^b\times\psi^c)(\phi_a\cdot\phi_b)(\phi_c\cdot\psi_3\times\psi_3)\\
        \frac{1}{2}(\psi_3\times\psi_3\cdot\phi^a)(\psi_a\cdot\phi^b\times\phi_b)(\psi^c\cdot\psi_c)\\
        (\psi_3\times\psi_3\cdot\psi^a)(\phi_a\cdot\psi^b)(\psi_b\cdot\phi^c\times\phi_c)\\
        -2(f\cdot\phi^a)(\phi_a\cdot\phi^b)(\phi_b\cdot\psi_3)(\psi^c\cdot\psi_c)\\
        -4(f\cdot\psi^a)(\phi_a\cdot\psi^b)(\phi_b\cdot\phi^c)(\phi_c\cdot\psi_3)\\
        -\frac{2}{3}(\phi^a\cdot\psi^b)(\psi^c\cdot\psi_3)(\psi_3\cdot\psi_a)(\phi_b\cdot\phi_c)\\
        -\frac{2}{3}(\phi^a\cdot\psi^b)(\psi^c\cdot\psi_3)(\psi_a\cdot\phi_c)(\phi_b\cdot\psi_3)\\
        -\frac{2}{3}(\psi^a\cdot\phi^b)(\psi^c\cdot\psi_3)(\psi_a\cdot\phi_c)(\phi_b\cdot\psi_3)
    \end{split}
\end{equation}

Order $M^{-2}$ ($O_{t^{28}}$):
\begin{equation}
    \begin{split}
        -2(\psi_3\times\psi_3\cdot\phi^a)(\psi_a\cdot\phi^b\cdot\phi_b)(\psi^c\cdot\phi_c\times\psi_3)\\
        4(f\cdot\phi^a)(\phi_a\cdot\phi^b)(\phi_b\cdot\psi_3)(\psi^c\cdot\phi_c\times\psi_3)\\
        2(f\cdot\phi^a)(\psi_a\cdot\phi^b\times\phi_b)(\phi^c\cdot\psi_3)(\phi_c\cdot\psi_3)\\
        4(f\cdot\psi^a)(\phi_a\cdot\phi^b\times\psi_3)(\phi_b\cdot\phi^c)(\phi_c\cdot\psi_3)\\
        4(f\cdot\phi^a\times\psi_3)(\phi_a\cdot\psi^b)(\phi_b\cdot\phi^c)(\phi_c\cdot\psi_3)\\
        -2(f\cdot\phi^a\times\phi_a)(f\cdot\phi^b)(\phi_b\cdot\phi^c)(\phi_c\cdot\psi_3)\\
        \frac{7}{3}(\phi^a\cdot\psi^b)(\psi^c\cdot\psi_3)(\phi_a\cdot\phi_c)(\phi_b\cdot\psi_3\times\psi_3)\\
        -\frac{2}{3}(\psi^a\cdot\phi^b)(\phi^c\cdot\psi_3\times\psi_3)(\psi_3\cdot\psi_a)(\phi_b\cdot\phi_c)\\
        \frac{2}{3}(\phi^a\cdot\psi^b)(\phi^c\cdot\psi_3\times\psi_3)(\phi_a\cdot\psi_b)(\phi_c\cdot\psi_3)\\
        -\frac{2}{3}(\phi^a\cdot\phi^b\times\psi_3)(\psi_3\cdot\psi^c)(\phi_a\cdot\phi_c)(\psi_3\times\psi_b)\\
        -(\psi_3\times\psi_3\cdot\phi^a)(\phi_a\cdot\phi^b)(\phi_b\cdot\psi_3)(\psi^c\cdot\psi_c)
    \end{split}
\end{equation}

%%%%%%%%%%%%%%%%%%%%%%%%%%%%%%%%%%%%%%%%%%%%%%%%%%%%%%%%%
%\vspace{6pt}

%%%%%%%%%%%%%%%%%%%%%%%%%%%%%%%%%%%%
%%%%%%%%%%%%%%%%%%%%%%%%%%%%%%%%%%%

\bibliographystyle{JHEP}
\bibliography{ref}

\end{document}